\documentclass[letterpaper,aps,twocolumn,showpacs,superscriptaddress,floats,prb]{revtex4}
\usepackage{amsmath}
\usepackage{graphicx,epsfig,dsfont}
\usepackage{amssymb}
\usepackage{ulem}
\usepackage{verbatim}
\usepackage{color}

\newcommand{\nn}{\nonumber}
\newcommand{\pdag}{{\phantom{\dagger}}}

\begin{document}

\title{Quantum phase transition in quantum wires controlled by an external gate}

\author{Tobias Meng}
\affiliation{Institut f\"ur Theoretische Physik, Universit\"at zu K\"oln,
Z\"ulpicher Str. 77, 50937 K\"oln, Germany}
\author{Mehul Dixit}
\affiliation{Department of Physics, The Ohio State University, Columbus, Ohio 43210, USA}
\author{Markus Garst}
\affiliation{Institut f\"ur Theoretische Physik, Universit\"at zu K\"oln,
Z\"ulpicher Str. 77, 50937 K\"oln, Germany}
\author{Julia S. Meyer}
\affiliation{SPSMS, UMR-E CEA / UJF-Grenoble 1, INAC, Grenoble, F-38054, France}

\begin{abstract}
We consider electrons in a quantum wire interacting via a long-range Coulomb potential screened by a nearby gate. We focus on the quantum phase transition from a strictly one-dimensional to a quasi-one-dimensional electron liquid, that is controlled by the dimensionless parameter $n x_0$, where $n$ is the electron density and $x_0$ is the characteristic length of the transverse confining potential. 
If this transition occurs in the low-density limit,
it can be understood as the deformation of the one-dimensional Wigner crystal to a zigzag
arrangement of the electrons described by an Ising order parameter.
The critical properties are governed by the charge degrees of freedom and the spin sector remains essentially decoupled. 
At large densities, on the other hand, 
the transition is triggered by the filling of a second one-dimensional subband of transverse quantization. 
Electrons at the bottom of the second subband
interact strongly due to the diverging density of states and become
impenetrable. We argue that this stabilizes the electron liquid as it suppresses pair-tunneling
processes between the subbands that would otherwise lead to an instability.
However, the impenetrable electrons in the second band are 
screened by the excitations of the first subband, so that the transition is identified as a Lifshitz transition of impenetrable polarons. We discuss the resulting phase diagram as a function of $n x_0$.
\end{abstract}

\date{\today}

\maketitle

\section{Introduction}

Electron correlations in one-dimensional (1D) systems are especially pronounced due to the restricted available phase space. Most prominently, the quasi-particle concept for the electron liquid, that accurately describes metals in higher dimensions, breaks down in 1D giving rise to a correlated Luttinger liquid with the concomitant spin-charge separation. 
At strong interactions, the formation of Wigner crystals and commensurate Mott insulating states are expected. 
Technological advances in the fabrication of quantum wires and carbon nanotube systems with a high tunability nowadays allows the controlled study of such strong 
correlation phenomena,\cite{Auslaender05, Steinberg07, Deshpande08,Jompol09, Deshpande09} see Ref.~\onlinecite{Deshpande10} for a recent review.

While transport and spectroscopic properties of quantum wires in the Luttinger liquid regime have been the subject of much attention over the years both experimentally\cite{exp-LL1, exp-LL2, exp-LL3, Auslaender02, Steinberg07,Jompol09} 
and theoretically,\cite{Kane92,Fisher97,GiamarchiBook} much less is known outside this regime where a description solely in terms of a Luttinger liquid becomes insufficient. This is, in particular, the case close to quantum phase transitions at which the number of Fermi points changes and, consequently, a Fermi energy $E_F$ vanishes. In the approximation of a non-interacting electron gas, these transitions are directly reflected as rounded steps in the conductance,\cite{vanWees88, Wharam88} $G(E_F,T)$, that sharpen up as the temperature $T$ decreases.
 The influence of electron-electron interactions close to these transitions is, however, only partially understood.\cite{Thomas96,Thomas00,Reilly01,Cronenwett02,Picciotto05,Rokhinson06,crook,Hew09,starikov,Matveev04,meir1,meir2,Balents00,Meyer07,Meyer09,Sitte09}
It is important to realize that the limits $E_F \to 0$ and $T\to 0$ in the approach to the quantum phase transition do not commute, and, in particular, the critical conductance $G(0,T)$ is governed by the underlying quantum critical point.\cite{Garst08} As a Fermi energy $E_F$ vanishes at the transition, one necessarily enters a regime where temperature $T$ is larger than characteristic energy scales of the electron liquid and, as a result, phenomena beyond a Luttinger liquid description become important. 
Close to the quantum phase transition, one can distinguish two characteristic energy scales, namely the Fermi energy $E_F$ for charge excitations and the spin exchange constant $J$ for spin excitations. The intermediate temperature range $E_F \gg  T \gg J$, where the charge sector still sustains plasmon excitations while the spin sector is already incoherent, has caught some attention recently.\cite{Matveev04,Cheianov04,Fiete04} It was pointed out by Matveev\cite{Matveev04} that in this spin-incoherent regime the conductance should qualitatively deviate from the prediction for the non-interacting electron gas, which might explain additional features in the conductance of quantum wires observed experimentally.\cite{Thomas00,Reilly01} 

In the present work, we consider the quantum phase transition from a strictly one-dimensional electron liquid to a quasi-one-dimensional state with the aim to identify the nature, i.e., the universality class, of the transition at $T=0$. In the absence of any interaction, the electrons form subbands whose separation is controlled by the confining potential. The quantum wire is one-dimensional if only the lowest subband is occupied and undergoes a transition to a quasi-one-dimensional state by filling the second subband. Without interaction, this is just a Lifshitz transition where the number of Fermi points increases from two to four. It is the purpose of this work to investigate how this transition is modified in the presence of Coulomb interaction between electrons that is screened by a nearby gate.

\begin{figure}
(a)
\includegraphics[width=0.13\textwidth]{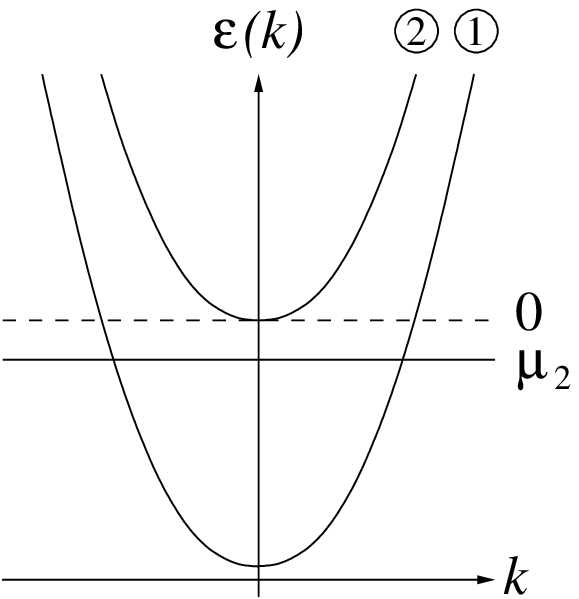}
\quad\quad
(b) 
\parbox[b]{5em}{\includegraphics[width=0.1\textwidth]{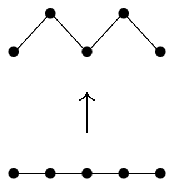}\\${}$}
\caption{Transition from a one-dimensional to a quasi-one-dimensional state. (a) At weak coupling, $n a_B \gg 1$, the transition is triggered by filling a second subband upon tuning the chemical potential $\mu_2$ through zero. (b) At strong coupling, $n a_B \ll 1$, the transition corresponds to the deformation of a 1D Wigner crystal to a zigzag configuration.}
\label{fig:transition}
\end{figure}

For spin-polarized electrons this question was already addressed in a series of works by some of the authors.\cite{Meyer07,Meyer09,Sitte09} 
It was found that in the limit of weak interactions, $n a_B \gg 1$, where $n$ is the one-dimensional electron density and $a_B$ is the Bohr radius, the transition is still a Lifshitz transition but in terms of polarons, i.e., dressed fermionic quasi-particles.  As the electrons start occupying the second subband, see Fig.~\ref{fig:transition}(a), they propagate with a very small velocity. The relatively fast Luttinger liquid fluctuations within the filled first subband can follow these electrons adiabatically thus dressing them with a cloud of plasmons. It was shown by Balents\cite{Balents00} that the residual couplings of the dressed fermions to the Luttinger liquid are irrelevant at the transition. A peculiarity of this transition concerns processes where pairs of electrons are transfered between the two subbands.\cite{Meyer07} Although such processes do not influence the nature of the  transition, they determine in fact the ground state for a finite density of polarons, i.e., pair-tunneling is {\it dangerously} irrelevant at the critical point. As polarons start populating the second subband, the system can gain energy by pair-tunneling and, as a result, a BCS-like gap opens up as a secondary effect. As a result of the dangerously irrelevant pair-tunneling operator, the Lifshitz transition thus separates two C1 phases in the notation of Ref.~\onlinecite{Balents96} that possess only a single gapless charge mode. 

As the interaction strength increases, the universality class of the transition, however, changes. For strong interactions, $n a_B \ll 1$, a one-dimensional Wigner crystal forms, and the transition corresponds to the splitting of the crystal into two rows with a zigzag arrangement of the electrons, see Fig.~\ref{fig:transition}(b). This transition 
can be described by an Ising order parameter. Correspondingly, the transition is now of Ising type and separates two C1 phases with a single gapless phonon mode. The residual coupling between the Ising critical degrees of freedom and the plasmon excitations leads to logarithmic corrections to Ising criticality, that were analyzed in Ref.~\onlinecite{Sitte09}. Although this coupling turns out to be marginally irrelevant, the renormalization group (RG) flow generates an enhanced SU(2) symmetry as the velocities of the plasmons and the Ising critical degrees of freedom approach each other in the low-energy limit. However, at the same time the mean velocity decreases resulting, e.g., in a divergence of the specific heat coefficient as a function of $T$. The multicritical point separating the Lifshitz transition at weak and the Ising transition at strong coupling has not been identified so far.

In this work, we extend the analysis of these previous works to electrons that are not spin polarized and investigate the influence of the spin degree of freedom on these transitions. One-dimensional two-subband systems of electrons have been investigated before by many authors.\cite{Varma85,Penc90,Fabrizio93,Balents96,Louis01,Japaridze07,Lai10} For a related theoretical study in the context of cold atomic systems see Ref.~\onlinecite{Rodriguez10}. Of particular interest in the present context are the works of Varma and Zawadowski,\cite{Varma85} that was motivated by the physics of fluctuating valence compounds, and of Balents and Fisher\cite{Balents96} that analyzes the phase diagram of the two-chain Hubbard model. We discuss the relation of our results to these earlier works in detail in the main text. 
The zigzag transition of the Wigner crystal was also studied 
in the context of ion traps\cite{ions1,ions2,ions3} as well as dipolar cold gases.\cite{dipolar}

Experimentally, the influence of the confining potential (and, thus, the interaction strength) on the transition from a one-dimensional to a quasi-one-dimensional state was investigated by Hew {\it et al.}\cite{Hew09} with the help of conductance measurements on a weakly-confined wire. The absence of the first quantized plateau in a range of confining potential strengths was interpreted as a signature related to the formation of a two-row Wigner crystal state.
 
The organization of this article is as follows. In section \ref{Model} we specify the model used in our study and construct a mean-field phase diagram. The transition for strong interactions, $n a_B \ll 1$, is discussed in section \ref{Wigner Crystal quantum wire} while the weak coupling regime, $n a_B \gg 1$, is addressed in section \ref{Multiband quantum wire}. We conclude in section \ref{Summary and Discussion} with a summary and discussion of our results.

\section{Model of the quantum wire}
\label{Model}

In the following, we specify the model of a quantum wire created by gating a two-dimensional electron gas (2DEG). The 2DEG forms at the interface (chosen to be in the $xy$-plane) between two different semiconductors which provide a steep confining potential in the perpendicular direction due to their band structure mismatch. Applying a bias voltage to a metallic split gate at a distance $d$ from the 2DEG leads to an additional confining potential $V_{\rm conf}(y)$ that restricts the motion of the electrons in the $y$-direction, thus creating a wire along the $x$-direction. In addition, the gate screens the Coulomb interaction between the electrons. 

The system is described by the model Hamiltonian 
\begin{align} \label{Model1}
\hat{H} = \hat{T} + \hat{V}_{\rm conf} + \hat{V}_{\rm int},
\end{align}
where the single-particle part comprises the kinetic energy, $\hat{T}$, and the confining potential, $\hat{V}_{\rm conf}$, whereas $\hat{V}_{\rm int}$ is the Coulomb interaction energy.
We work in units $\hbar = 1$, $k_B = 1$, and $4\pi\epsilon_0 = 1$. 

Assuming a parabolic confining potential characterized by the frequency $\Omega$, the single-particle terms read
\begin{align} \label{SingleParticleHam}
\hat{T} = \sum_i \frac{\hat{p}^2_i}{2m},\quad
\hat{V}_{\rm conf} = \frac{1}{2} m \Omega^2 \sum_i \hat{y}^2_i,
\end{align}
where the index $i$ labels the electrons and $m$ is the effective electron mass. The interaction between electrons is given by
\begin{align}
\hat{V}_{\rm int} = \frac{1}{2} \sum_{i \neq j} U(\hat{\bf r}_i - \hat{\bf r}_j) 
\end{align}
with the screened Coulomb potential
\begin{align} \label{Interaction1}
U({\bf r}) = \frac{e^2}{\epsilon} \Big(\frac{1}{|{\bf r}|} - \frac{1}{\sqrt{{\bf r}^2 +(2d)^2} } \Big),
\end{align}
where $e$ is the electron charge and $\epsilon$ is the dielectric constant of the host material. The mirror charge in the gate ensures that the potential falls off as a dipole field $U({\bf r}) \approx 2 e^2 d^2/(\epsilon |{\bf r}|^3)$ for large distances, $r \gg d$.

\subsection{Mean-field phase diagram}

Before analyzing the Hamiltonian \eqref{Model1}, let us discuss its qualitative features.

\begin{figure}
\includegraphics[width=0.4\textwidth]{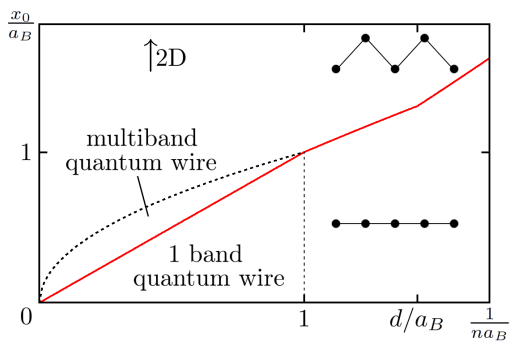} 
\caption{Mean-field phase diagram of the Hamiltonian \eqref{Model1} as a function of the inverse one-dimensional electron density $n^{-1}$ and the oscillator length $x_0=1/\sqrt{m \Omega}$, both measured in units of the  Bohr radius $a_B = \epsilon/(e^2 m)$. As $x_0$ increases, a transition occurs at the (red) solid line from a one-dimensional to a quasi-one-dimensional state. At low densities $1/(n a_B) \gg 1$, this transition corresponds to the deformation of a 1D Wigner crystal into a zigzag configuration. At high densities, $1/(n a_B)\ll 1$, the transition is triggered by the filling of a second subband. The thick dotted line indicates where the interaction energy equals the subband separation so that the band picture ceases to be well-defined. For large $x_0$, the two-dimensional limit is approached. For a derivation of the various lines and regimes, see text.     
}
\label{fig1}
\end{figure}

\subsubsection{Characteristic length scales} 

The Hamiltonian \eqref{Model1} is characterized by four characteristic length scales. {\it (i)} The mean particle spacing is given by the inverse one-dimensional density, $1/n$, i.e., the number of particles in the quantum wire per unit length. {\it (ii)} The extension of the system in the lateral direction is quantified by the oscillator length, $x_0 = 1/\sqrt{m \Omega}$. Finally, the electron-electron interaction is characterized by {\it (iii)} the Bohr radius, $a_B = \epsilon/(e^2 m)$, and {\it (iv)} the distance to the gate, $d$, specifying the strength and range of the interaction potential \eqref{Interaction1}, respectively. In the following, we assume that the Bohr radius is much smaller than the distance to the gate, $a_B \ll d$, as is usually the case for quantum wires; for example, for GaAs based systems one can estimate $a_B \simeq 10\,$nm and $d \gtrsim 100\,$nm.\cite{Meyer09}

To understand the interplay between the different length or energy scales, a mean-field phase diagram of the model as a function of the length scales $1/n$, $x_0$, $a_B$, and $d$ can be constructed by minimizing the dominant terms in the Hamiltonian in various regimes and comparing the associated ground state energy scales. In particular, one can distinguish the cases where either the single-particle part of the Hamiltonian dominates over the interaction energy $\hat{V}_{\rm int}$ (Sec.~\ref{sssec-single}) or vice versa (Sec.~\ref{sssec-interaction}). By comparing the Fermi energy, $E_F\sim n^2/m$, which is the characteristic energy for single-particle physics, with the typical interaction energy at distances of order of the mean particle spacing, $U(1/n)\sim ne^2/\epsilon$, one finds that the two cases are distinguished by the ratio of the mean particle spacing and the Bohr radius. For high densities, $na_B>1$, the single-particle energy dominates whereas for low densities, $na_B<1$, the interaction energy dominates.

\subsubsection{Single-particle limit: multi-subband quantum wire}
\label{sssec-single} 

In the high density limit, $na_B\gg1$, interactions are weak.
Thus, it is appropriate to first chose the eigenbasis of the single-particle part of the Hamiltonian consisting of $\hat{T} + \hat{V}_{\rm conf}$, see Eqs.~\eqref{SingleParticleHam}, and then analyze the effect of interactions in that basis. The single-particle eigenbasis is given by product-wavefunctions of traveling waves along the wire and the oscillator eigenfunctions in the transverse directions. The system is one-dimensional (1D) as long as only the lowest oscillator level is occupied. By comparing the Fermi energy with the oscillator frequency $\Omega$, one obtains the condition 
\begin{align}
x_0 < \frac{1}{n},
\end{align} 
that is shown as a solid line in Fig.~\ref{fig1}. The transition from a single- to a two-subband quantum wire that is at the focus of this work occurs for $n x_0 \sim 1$. Upon further increasing the density $n$ or relaxing the confining potential, i.e., increasing $x_0$, more and more subbands are populated.

The oscillator levels are well defined as long as the oscillator frequency $\Omega$ is larger than the typical Coulomb energy $U(1/n)$, which translates to the condition $x_0< \sqrt{a_B/n}$, see dotted line in Fig.~\ref{fig1}. For even larger oscillator length $x_0$ one crosses over into a two-dimensional regime where the subbands are washed out. 

\subsubsection{Interaction dominated regime: Wigner crystal}
\label{sssec-interaction} 

For small densities, $n a_B \ll 1$, interactions dominate over the kinetic energy of the particles, and the system behaves almost classically. In that case, the potential energy, $\hat{V}_{\rm conf} + \hat{V}_{\rm int}$, should be minimized first, resulting in a Wigner crystal state of electrons.\cite{Wigner, Schulz93} As a function of oscillator length $x_0$, there is a competition between the interaction and the confining potential. At strong confinements, $x_0 \to 0$, a one-dimensional Wigner crystal is the ground state of the potential energy $\hat{V}_{\rm conf} + \hat{V}_{\rm int}$. If the confinement is relaxed, a transition to a zigzag Wigner crystal occurs, followed by subsequent transitions to more two-dimensional multirow structures.\cite{zigzag1, zigzag2, zigzag3, Meyer07, Meyer09, CortesHuerto10} Comparing $\hat{V}_{\rm conf}$ with the energy gain obtained by relaxing the Wigner crystal from a strictly one-dimensional arrangement, $U({\bf r}=(1/n,0)) - U({\bf r}=(1/n,y))$, one arrives at the stability criterion for the one-dimensional Wigner crystal state, 
\begin{align} \label{WC-criterion}
x_0 < \left(\frac{a_B}{n^3}\right)^{1/4},
\end{align}
indicated by the solid line in Fig.~\ref{fig1}. Note that, at very low densities, $n < 1/d$, the screening of the interaction becomes important and modifies the transition line between the one-dimensional and the zigzag crystal. 
At even lower densities, $n<a_B/d^2$, the Wigner crystal melts (not shown in Fig.~\ref{fig1}).

\subsection{Outline}
The focus of this work is the evolution of the quantum wire from the strictly one-dimensional limit to a quasi-one-dimensional state. At fixed Bohr radius $a_B$ and distance to the gate $d$, a transition to a quasi-one-dimensional state occurs upon increasing oscillator length $x_0$ or density $n$, see Fig.~\ref{fig1}. This transition is the subject of the present paper with a focus on the interplay of charge and spin degrees of freedom. In the strongly interacting Wigner crystal regime, the increase of $n$ or $x_0$ leads to a transition from a 1D to a zigzag Wigner crystal. Charge and spin degrees are governed by very different energy scales so that, as a starting point, 
they can be treated separately as done in Refs.~\onlinecite{Meyer07,Sitte09} (charge) and \onlinecite{Klironomos} (spin). The question as to how the spin sector affects the Wigner crystal transition is addressed in Sec.~\ref{Wigner Crystal quantum wire}. In the weakly interacting regime, the increase of $n$ or $x_0$ results in the subsequent population of subbands. We focus on the transition where the second subband starts to get filled and analyze the effect of interactions. While a single subband can be described as a Luttinger liquid displaying spin-charge separation, the interactions between the subbands couple spin and charge degrees of freedom. This leads to interesting modifications as compared to the spin-polarized case,\cite{Meyer07,Sitte09} that will be discussed in Sec.~\ref{Multiband quantum wire}.

\section{Wigner Crystal quantum wire}
\label{Wigner Crystal quantum wire}

At low electron densities, $n a_B \ll 1$, the interaction energy dominates over the Fermi energy, see Fig.~\ref{fig1}, and the classical ground state is a good starting point to describe the physics of the system. Minimizing the potential energy $\hat{V}_{\rm conf} + \hat{V}_{\rm int}$ of Eq.~\eqref{Model1} one arrives at a Wigner crystal state for the electrons.\cite{Meyer09}  At low electron densities or strong confinement such that Eq.~\eqref{WC-criterion} is obeyed, this Wigner crystal corresponds to a one-dimensional arrangements of electrons with equilibrium position $(x^0_j,y^0_j) = (j\, a,0)$ of the $j^{th}$ electron, where $a = n^{-1}$ is the lattice spacing. If the confinement is relaxed and the oscillator length $x_0$ increases, a transition from a one-dimensional arrangement of electrons to a zigzag Wigner crystal takes place, see Fig.~\ref{fig:transition}(b). In the following, we consider the properties of this quantum phase transition. 

\subsection{Charge sector}

The excitations of the one-dimensional Wigner crystal correspond to fluctuations of the electrons around their equilibrium positions, 
\begin{align}
(u_{x j}, u_{y j}) = \Big(\frac{x_j - x_j^0}{a}, \frac{y_j-y_j^0}{a}\Big) 
= \Big(\frac{x_j}{a} -j,\frac{y_j}{a}\Big),
\end{align}
which are the phonon modes. Deep in the 1D regime, the only low-energy mode is the Goldstone mode $u^\parallel_0$ corresponding to a translation of the Wigner crystal along the wire direction, $(u_{x j}, u_{y j}) = u^\parallel_0 (1,0)$. As the confining potential is weakened and the quantum phase transition is approached, there appears another low-energy mode $u^\perp_\pi$. It corresponds to the alternating oscillation of electrons in the direction transverse to the wire, $(u_{x j}, u_{y j}) = u^\perp_\pi (0, (-1)^j)$. The remaining two phonon modes, the longitudinal out-of-phase phonon mode $u^\parallel_\pi ((-1)^j,0)$ and the transversal phonon $u^\perp_0 (0,1)$, have gaps determined by the Coulomb repulsion and the confining potential, respectively. Close to the transition both these gaps are of the order $\Omega$.

The Lagrangian density $\mathcal{L} = \mathcal{L}_{\parallel 0} + \mathcal{L}_{\perp \pi} + \mathcal{L}_{\mathrm{int}}$ governing these low-energy excitations in the continuum limit is given by
\begin{subequations}
\label{CriticalModelZigzag}
\begin{align}
\mathcal{L}_{\parallel0} &= 
\frac{m}{2 n}
\left[ {(\partial_\tau u_0^\parallel)}^2+v_{\parallel0}^2 \left(\partial_x u_0^\parallel\right)^2\right],\\
\mathcal{L}_{\perp\pi} &= 
\frac{m}{2 n} \left[{(\partial_\tau u_\pi^\perp)}^2 + v_{\perp\pi}^2 \left(\partial_x u_\pi^\perp\right)^2+r\,(u_\pi^\perp)^2+s\,(u_\pi^\perp)^4\right],\label{L_charge-perp}\\
\mathcal{L}_{\rm int} &= \lambda
\left(\partial_x u_0^\parallel\right) {u_\pi^\perp}^2,\label{L_charge-int}
\end{align}
\end{subequations}
where $\tau$ is the imaginary time. The longitudinal velocity $v_{\parallel 0}$ is determined by the compressibility, $m v_{\parallel0}^2/n = \partial^2 E_0/\partial n^2$, where $E_0$ is the ground state energy; in the limit $d^{-1} \ll n\ll a_B^{-1}$, it evaluates to $v_{\parallel0}^2 = 2 n/(m^2 a_B) \ln(8 n d)$.\cite{Meyer09} The control parameter $r$ in (\ref{L_charge-perp}) changes sign as function of $x_0$, namely 
$r=1/m^2 ({x_0}^{-4} - {x_{0c}}^{-4})$ with $x_{0c} = (2\epsilon/(7\zeta(3)me^2n^3))^{1/4}$. The transverse velocity $v_{\perp\pi}$ and coefficient of the quartic term $s$ read $v_{\perp\pi}^2 = n/(m^2 a_B) \ln(2)$ and $s = 93\zeta(5)n^4/(8m^2a_B)$, respectively. Finally, the coupling between the two modes is given by $\lambda = 21 \zeta(3) e^2 n/(4\epsilon)$. 

Away from the transition $r>0$, the transverse mode $u_\pi^\perp$ corresponds to an optical phonon with minimal frequency $\propto \sqrt{r}$. 
With increasing oscillator strength $x_0$, this gap decreases and vanishes at a critical value $x_{0c}$, signaling the instability of the one-dimensional Wigner crystal. For $r<0$ the phonon field $u_{\pi}^\perp$ condenses with a finite expectation value corresponding to a zigzag Wigner crystal phase, see Fig.~\ref{fig:transition}(b). For $r=0$, the Lagrangian $\mathcal{L}_{\perp\pi}$ corresponds to a critical one-dimensional Ising model $(s>0)$. Neglecting the coupling between modes, $\lambda$, the transition from the 1D to the zigzag Wigner crystal is, thus, in the Ising universality class and the transverse mode has a gap which scales linearly with the distance from the transition point, 
$\Delta_\pi^\perp= r$. If the Wigner crystal was pinned and the positions of particles along the wire were fixed, the transition would break the reflection symmetry in the confining plane. However, the presence of the mode $u_\parallel^0$, i.e., the fact that the crystal may deform in the longitudinal direction, makes the zigzag order non-local.\cite{Ruhmann}

It is well-known that the transverse field Ising model can be alternatively represented in terms of a fermionic degree of freedom $\Psi$ with 
Lagrangian density\cite{Pfeuty}
\begin{align} \label{FermionL}
\mathcal{L}^{\rm ferm}_{\perp\pi} = 
\Psi^\dagger \partial_\tau \Psi^\pdag + \frac{v_{\perp \pi}}{2} \left(\Psi \partial_x \Psi + {\rm h.c.}\right) + r \Psi^\dagger \Psi^\pdag.
\end{align}
The longitudinal plasmon $u^0_{\parallel}$ couples to the most relevant operator $(u_\pi^\perp)^2 \sim \Psi^\dagger \Psi$ of the Ising model so that the interaction term (\ref{L_charge-int}) can be rewritten in the fermionic formulation as 
\begin{align} \label{CouplingL}
\mathcal{L}^{\rm ferm}_{\rm int} &= 
\lambda
\left(\partial_x u_0^\parallel\right) \Psi^\dagger \Psi^\pdag,
\end{align}
where, for simplicity of notation, we suppressed in Eqs.~\eqref{FermionL} and \eqref{CouplingL} renormalizations of coupling constants.

The model $\mathcal{L}_{\parallel0} + \mathcal{L}^{\rm ferm}_{\perp\pi} + \mathcal{L}^{\rm ferm}_{\rm int}$ and its critical properties were analyzed and discussed in Ref.~\onlinecite{Sitte09}. It was found that the critical renormalization group flow of the model parameters depends on the ratio of velocities, $v_{\perp\pi}/v_{\parallel0}$. If $v_{\perp\pi} < v_{\parallel0}$, which is the case for quantum wires (see above), the interaction $\lambda$ is marginally irrelevant and decreases with decreasing energy. At the same time the ratio of velocities $v_{\perp\pi}/v_{\parallel0}$ approaches one. The critical fixed point is, thus, characterized by an enhanced SU(2) symmetry. However, a peculiarity of the RG flow is that the velocity $v_{\parallel0}$ itself vanishes in the low-energy limit resulting, e.g., in a diverging specific heat coefficient at the critical point. (In the opposite limit, $v_{\perp\pi} > v_{\parallel0}$, run-away RG flow was found.)

In the following, we address the question whether these critical properties are modified in the presence of a coupling to the spin degrees of freedom.

\subsection{Coupling to the spin sector}

Spin interactions in the one-dimensional Wigner crystal are described by the anti-ferromagnetic Heisenberg model with nearest-neighbor interactions. In the zigzag Wigner crystal, next-nearest neighbor interactions as well as ring exchange processes become important and lead to rich spin physics.\cite{Klironomos} 
However, these additional interactions become important only once the lateral extent of the crystal is sufficiently large. Close to the transition, they are negligible, and the spin interactions are still described by the Heisenberg Hamiltonian,
\begin{equation}
\mathcal{H}_s = J \sum_j \vec{S}_j \cdot \vec{S}_{j+1} ,
\end{equation}
where the coupling constant $J$ is exponentially small in $1/(na_B)$.\cite{Matveev04}

Due to the exponential dependence of the spin interactions on the inter-particle distance, fluctuations of the electron positions immediately result in a modulation of $J$ and, therefore, give rise to a magnetoelastic coupling between the spins and 
the charge modes. Specifically, the interaction energy $J_j$ between electron $j$ and $j+1$ depends on the position of both electrons,\cite{footnote1} $J_j=J({\bf r}_j,{\bf r}_{j+1})\simeq J(|{\bf r}_{j+1}-{\bf r}_j|)$. In order to investigate the coupling between spin and charge modes perturbatively, we expand the interaction energy $J_j$ in small fluctuations around the equilibrium positions of the electrons using $|{\bf r}_{j+1}-{\bf r}_j|=a\sqrt{(1+u_{xj+1}-u_{xj})^2+(u_{yj+1}-u_{yj})^2}$. Thus,
\begin{eqnarray}
J_j\simeq J(a)+a J'(a) \left(u_{xj+1}\!-\!u_{xj}+\frac12(u_{yj+1}\!-\!u_{yj})^2\right).
\nonumber
\end{eqnarray}
Note that the expansion of the coupling $J$ in the longitudinal fluctuations $u_{xj}$ starts in linear order, yielding
\begin{equation} \label{LongCoupling}
\mathcal{H}^\parallel_{s c} = -g_\parallel  \sum_j (u_{xj+1} - u_{xj}) \vec{S}_j\cdot \vec{S}_{j+1},
\end{equation}
with $g_\parallel = - a J'(a)$. By contrast, due to the symmetry of the one-dimensional Wigner crystal, the expansion in the transverse fluctuations $u_{y j}$ begins only in second order, i.e.,
\begin{align} \label{TransCoupling}
\mathcal{H}^\perp_{s c} &= -g_\perp \sum_j (u_{yj+1} - u_{yj})^2 \vec{S}_j \cdot \vec{S}_{j+1},
\end{align}
with $g_\perp = - a J'(a)/2$.

The linear coupling to the longitudinal mode (\ref{LongCoupling}) is familiar from the spin-Peierls problem.\cite{GiamarchiBook} In particular, the mode $u^\parallel_\pi$ with momentum $q=\pi$, $u_{x j} = u^\parallel_\pi (-1)^j$, couples to the staggered part of $\vec{S}_j\cdot \vec{S}_{j+1}$,
\begin{equation} \label{LongCoupling_pi}
\mathcal{H}^\parallel_{s c} \approx -2 g_\parallel  \sum_j u^\parallel_\pi(-1)^j \vec{S}_j\cdot \vec{S}_{j+1}.
\end{equation}
If the $u^\parallel_\pi$ mode was sufficiently soft, this term would lead to a spin-Peierls transition. The crystal distorts such that the mode $u^\parallel_\pi$ assumes a non-vanishing expectation value giving rise to an alternation of weak and strong bonds, $J\pm\delta J$. The system then gains magnetic energy by forming singlets on the strong bonds.\cite{Cross79}
In our case, however, the magnetic energy is exponentially small such that it never can compete with the charge gap of the $u^\parallel_\pi$ mode, that is on the order of $(n/m)\sqrt{n/a_B}\ln(nd)$. We can, thus, conclude that the interaction \eqref{LongCoupling} of the spin degrees of freedom with the longitudinal modes does not influence the critical properties of the charge sector. 

We now turn to the coupling to the transverse mode \eqref{TransCoupling}. The most singular contribution is attributed to the critical $u_\pi^\perp$ mode. Substituting $u_{y j} = u_\pi^\perp (-1)^j$, we obtain
\begin{align} \label{TransCouplingSoft}
\mathcal{H}^\perp_{s c} 
&\approx 
-4 g_\perp \sum_j (u_\pi^\perp)^2  \vec{S}_j \cdot \vec{S}_{j+1}.
\end{align}
It turns out, however, that this interaction is also irrelevant as far as the critical properties of the Ising transition is concerned. This conclusion  follows from a straightforward power counting analysis of the Ising operator, $(u_\pi^\perp)^2 \sim \Psi^\dagger \Psi$, and the non-staggered spin-spin operator, $\vec{S}_j \cdot \vec{S}_{j+1}$. 

Both magnetoelastic couplings, $g_{\parallel}$ and $g_{\perp}$, therefore do not modify the transition to the zigzag Wigner crystal described by the model \eqref{CriticalModelZigzag}. The transition happens only in the charge sector and the spin sector acts as a spectator. The spin and charge degrees thus remain essentially decoupled at the transition and the analysis of Ref.~\onlinecite{Sitte09} still applies. 
We now turn to the analysis of the transition in the weak-coupling limit $n a_B \gg 1$.

\section{Multiband quantum wire}
\label{Multiband quantum wire}

At large densities, $n a_B > 1$, there exists a regime at sufficiently small oscillator length, $x_0 < \sqrt{a_B/n}$, where the quantum wire contains well-defined single-particle subbands, see Fig.~\ref{fig1}. In this regime, we can use the single-particle basis of the Hamiltonian \eqref{Model1}, deriving from the kinetic energy and the confining potential, $\hat{T} + \hat{V}_{\rm conf}$, and expand the interaction $\hat{V}_{\rm int}$ in that basis. 

In second quantized form, the resulting Hamiltonian reads
\begin{align} \label{HamPert1}
&\mathcal{H} = \sum_{n, k, \sigma} 
\left(\frac{k^2}{2 m} - \mu_n\right) c^\dagger_{n k \sigma} c^\pdag_{n k\sigma}
\\\nn&+
\frac{1}{2}\mathop{\sum_{n_1,n_2,n_3,n_4}}_{k,k',q;\sigma, \sigma'}
U_{n_1 n_2 n_3 n_4}(q) c^\dagger_{n_1 k+q \sigma} c^\dagger_{n_2 k'-q \sigma'} 
c^\pdag_{n_3 k' \sigma'} c^\pdag_{n_4 k \sigma},
\end{align}
where the electron operators $c_{n k \sigma}$ destroy an electron with subband index $n$ (i.e., the quantum number of the harmonic oscillator defined by $\hat V_{\rm conf}$), momentum $k$ in $x$-direction along the wire, and spin $\sigma= \uparrow,\downarrow$. Two consecutive chemical potentials differ by the oscillator frequency, $\mu_{n} - \mu_{n+1} = \Omega$. 

The electrons interact with an interaction amplitude $U$ that depends on the transfered (longitudinal) momentum $q$ and the subband indices $n_i$. Its value is given by matrix elements of the screened Coulomb interaction in the basis of oscillator wavefunctions,
\begin{align} \label{Interaction2}
U_{n_1 n_2 n_3 n_4}(q) = \int \frac{dq_y}{2\pi}\, U(q,q_y) \Gamma_{n_1 n_2 n_3 n_4}(q_y),
\end{align}
where the Fourier transform of the interaction is given by 
\begin{align} \label{Interaction3}
U(q,q_y) &= \int d{\bf r}\; e^{-i{\bf qr}} \,U({\bf r})
\\\nn&= \frac{e^2}{\epsilon} \frac{2\pi}{\sqrt{q^2 + {q_y}^2}} \left(1 - e^{-2 d \sqrt{q^2 +  {q_y}^2}}\right)
\end{align}
and the matrix elements read
\begin{align}
\lefteqn{\Gamma_{n_1 n_2 n_3 n_4}(q_y) =}
\\&\nn
\int dy_1 dy_2\; e^{i q_y(y_1 - y_2)}
\phi^*_{n_1}(y_1)
\phi^*_{n_2}(y_2)\phi_{n_3}(y_2)\phi_{n_4}(y_1)
\end{align}
with the $n^{\rm th}$ 1D oscillator wavefunctions $\phi_n(y)$. Restricting ourselves to the low-energy properties of the system, only interaction matrix elements $U_{n_1n_2n_3n_4}$ where the indices $n_i$ are pair-wise equal will appear.

In the following, we focus on the situation where the first subband is filled, $\mu_1 > 0$, and the density of electrons in the second subband is very dilute, $|\mu_2| \ll \mu_1$, i.e., the second subband is close to the quantum phase transition occurring at $\mu_2 =0$, see Fig.~\ref{fig:transition}(a). In this regime, we can neglect all higher subbands so that the band index is restricted to $n = 1,2$. The corresponding oscillator eigenfunctions are
\begin{subequations}
\begin{align}
\phi_{1}(y) &= \left(\frac{m\Omega}{\pi}\right)^{1/4} e^{-\frac{1}{2} m \Omega y^2},
\\
\phi_{2}(y) &= \sqrt{2 m \Omega}\left(\frac{m\Omega}{\pi}\right)^{1/4} y \;e^{-\frac{1}{2} m \Omega y^2}.
\end{align}
\end{subequations}
The reduced Hamiltonian for the two subband system is, thus,
\begin{equation} \label{HamPert2}
\mathcal{H} = \mathcal{H}_1 + \mathcal{H}_2 + \mathcal{H}_{12},
\end{equation}
where $\mathcal{H}_i$ represent the two (interacting) subbands, $i =1,2$, and $\mathcal{H}_{12}$ captures the inter-subband interactions. The condition $x_0<\sqrt{a_B/n}$, that is fulfilled below the dotted line in Fig.~\ref{fig1}, implies that the dimensionless interaction $\nu_1U_{n_1 n_2 n_3 n_4}(q) \ll 1$ is small, where $\nu_1 = (2/\pi) \sqrt{m/(2 \mu_1)}$ is the density of states of the filled first subband.

\subsection{Low-energy limit of the Hamiltonian}
\label{ssec-H12}

In the following, we discuss the different parts of the two-subband Hamiltonian \eqref{HamPert2} separately  before turning to the analysis of the full Hamiltonian in subsequent sections.

\subsubsection{First subband}
\label{sssec-H1}

The Hamiltonian $\mathcal{H}_{1}$ in Eq.~\eqref{HamPert2} represents the electrons in the first subband at finite $\mu_1 > 0$, interacting via the potential $U_{1111}(q)$. At low temperatures $T \ll \mu_1$, we can approximate the first subband by a Luttinger liquid, linearizing the dispersion around the two Fermi points at $\pm k_{F1}=\pm\sqrt{2m\mu_1}$. 
Introducing right- and left-moving fields, denoted $R$ and $L$, respectively,
\begin{equation}
c_{1 \sigma}(x) = e^{i k_{F1} x} R_{1\sigma}(x) + e^{-i k_{F1} x} L_{1\sigma}(x),\label{eq-RL}  
\end{equation}
the Hamiltonian $\mathcal{H}_1$ takes the standard form 
\begin{eqnarray} \label{Band1Ham1}
\mathcal{H}_1 &=& \int dx \Big[
-iv_{F1}\sum_\sigma
\left( R^\dagger_{1 \sigma}\partial_x R^\pdag_{1\sigma} -L^\dagger_{1 \sigma}\partial_xL^\pdag_{1\sigma}\right)\quad
\\\nn
&&\qquad+ g_{1c}\rho_{1R}\rho_{1L} - g_{1s}\vec{S}_{1R}.\vec{S}_{1L}\Big].
\end{eqnarray}
Here the particle and spin densities of right ($r = R$) and left ($r = L$) moving electrons are given by
\begin{align}
\label{eq-rhoS}
\rho_{1r} &= \sum_{\sigma}r^{\dag}_{1\sigma}r^\pdag_{1\sigma},\qquad
\vec{S}_{1r} &= \frac{1}{2}\sum_{\sigma,\sigma'}r^{\dag}_{1\sigma}\vec{\sigma}_{\sigma,\sigma'}r^\pdag_{1\sigma'},
\end{align}
where $\sigma_i$ are Pauli matrices. 
The interaction constants describing scattering processes in the vicinity of the Fermi surface are given by 
\begin{align}
\label{eq-int1}
g_{1c} = U_{1111}(0) - \frac{U_{1111}(2k_{F1})}{2},\quad
g_{1s} = 2U_{1111}(2k_{F1}).
\end{align}
Their magnitude is evaluated and discussed in Sec.~\ref{subsub:CouplingConst}. Here chiral interactions which only renormalize the velocities of the charge and spin modes are neglected. The Hamiltonian ${\cal H}_1$ describes decoupled charge and spin modes, and has been 
intensively studied.\cite{GiamarchiBook} 
The charge mode is gapless whereas the fate of the spin mode depends on the sign of $g_{1s}$. If $g_{1s}$ 
were negative, the spin mode would acquire a gap. In the present case, however, $g_{1s}$ is positive, see Eq.~\eqref{eq-int1}, and, therefore, for a single subband quantum wire, the spin sector 
remains gapless.

\subsubsection{Second subband}
\label{sssec-H2}
 
The electrons in the second subband are close to the band bottom, $\mu_2 \approx 0$, and a clear distinction of left- and right-movers and subsequent linearization of the spectrum is in general not possible. Furthermore, due to the low particle-density at the band bottom only interactions between electrons of different spin polarizations are of importance. Local interactions among electrons with the same spin are suppressed due to the Pauli principle. The residual interaction contains gradients of the electron fields; 
it is irrelevant in the RG sense at the critical point $\mu_2 = 0$ and, therefore, will be neglected. The resulting Hamiltonian for the second subband reads
\begin{align} \label{Band2Ham}
\mathcal{H}_2 = \int dx 
\Big[
&
\sum_\sigma c^\dagger_{2 \sigma}(x) \left(
-\frac{\partial_x^2}{2 m} - \mu_2 \right) c^\pdag_{2\sigma}(x) 
\\\nn&+
V\, c^\dagger_{2 \uparrow}(x) c^\dagger_{2 \downarrow}(x) 
c^\pdag_{2 \downarrow}(x) c^\pdag_{2 \uparrow}(x)
\Big],
\end{align} 
where
\begin{align}
V = U_{2222}(0).
\end{align}
For small $\mu_2$, the Hamiltonian ${\cal H}_2$ represents a strongly interacting system even for infinitesimally small interaction $V$. This is best seen for negative chemical potential, $\mu_2 < 0$, where the ground state of Eq.~\eqref{Band2Ham} is empty of particles. In this case, the retarded two-particle Green function, 
\begin{align}
\lefteqn{\mathcal{D}_{2}(x-x',t-t') = }
\\\nn&
- i \theta(t-t')\langle [c^\pdag_{2\uparrow}(x,t) c^\pdag_{2\downarrow}(x,t),
c^\dagger_{2\downarrow}(x',t') c^\dagger_{2\uparrow}(x',t')] \rangle,
\end{align}
can be straightforwardly evaluated at $T=0$ because it only requires the solution of a two-particle problem. The bare retarded two-particle function, $\mathcal{D}^{(0)}_2$, in the absence of interactions has the form
\begin{align} \label{G2Bare1}
\mathcal{D}^{(0)}_2(k,\omega) &= 
- i \frac{\sqrt{m}}{2 \sqrt{\omega - \frac{k^2}{4 m} + 2 \mu_2 + i 0}}=- \frac i2 \sqrt{\frac{m}{\varepsilon+ i 0}},
\end{align}
where $\varepsilon=\omega - \frac{k^2}{4 m} + 2 \mu_2$ is the distance to the two-particle mass shell. 

\begin{figure}
(a) \parbox[b]{0.4\textwidth}{\includegraphics[width=0.4\textwidth]{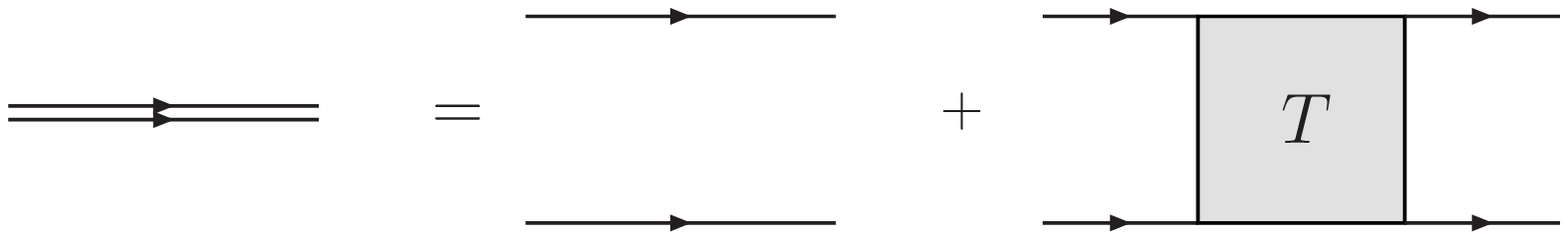}}\\[0.5em]
(b) \parbox[b]{0.4\textwidth}{\includegraphics[width=0.35\textwidth]{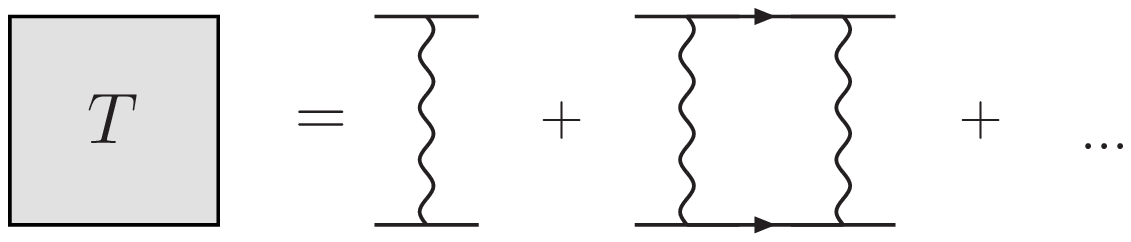}}
\caption{(a) Two-particle Green function \eqref{G2} of two electrons with opposite spin polarizations in the second subband. (b) The repeated scattering of the two electrons leads to the T-matrix, Eq.~\eqref{Tmatrix}, where the wiggly line represents the interaction $V$. 
}
\label{fig2}
\end{figure}

In the presence of the interaction $V$, the two particles repeatedly scatter off each other. The exact retarded two-particle Green function is then obtained by summing the interaction ladder in the particle-particle channel, see Fig.~\ref{fig2}, resulting in
\begin{align}\label{G2}
\mathcal{D}^{-1}_2(k,\omega) = {\mathcal{D}^{(0)-1}_2}(k,\omega) - V.
\end{align}
Alternatively, this Green function can be expressed in terms of the two-particle T-matrix, $\mathcal{D}_2(k,\omega) = \mathcal{D}^{(0)}_2(k,\omega) + \mathcal{D}^{(0)}_2(k,\omega) \mathcal{T}(k,\omega)\mathcal{D}^{(0)}_2(k,\omega)$, where 
\begin{align} \label{Tmatrix}
\mathcal{T}(k,\omega) = \frac{V}{1-V\, \mathcal{D}^{(0)}_2(k,\omega)}.
\end{align}
Due to the inverse square-root singularity of $\mathcal{D}^{(0)}_2(k,\omega)$, the T-matrix takes a universal form as the two-particle mass shell is approached, $\varepsilon\to0$,
\begin{align} \label{UnitaryLimit}
\mathcal{T}(k,\omega) \to - 2i \sqrt{\frac{\varepsilon+ i 0}{m}}.
\end{align}

This corresponds to unitary scattering with phase shift $\delta = \pi/2$. In particular, note that the same limit obtains from Eq.~\eqref{Tmatrix} for infinitely strong repulsion $V \to \infty$. As the low-energy limit $k,\omega\to0$ close to criticality $\mu_2 = 0$ coincides with the limit $\varepsilon \to 0$, the second subband is populated by a strongly interacting electron gas (irrespective of the value of $V$), where two electrons with opposite spin polarization cannot occupy the same state. As a consequence, the two-particle wavefunction has not only nodes for electrons with the same spin as required by the Pauli principle, but also for electrons with opposite spin polarizations. Such a strongly interacting gas has been dubbed an {\it impenetrable electron gas}.\cite{Ogata90,EsslerBook,Takahashi71,Goehmann98-2} The impenetrable electron gas is the fixed point theory that describes the quantum phase transition in the second subband at $\mu_2 = 0$ in the absence  of inter-subband interactions.

At positive $\mu_2$, the ground state of the system is a Luttinger liquid. For small $0< \mu_2 < E_p = m V^2$, the strong correlations of the emerging impenetrability of electrons are reflected in largely different energy scales governing the charge and spin sector. While the characteristic energy scale for charge excitations is the Fermi energy $\mu_2$, the one for spin excitations, $J$, arises from corrections to the unitary limit \eqref{UnitaryLimit} and is given by\cite{Cheianov04} $J \sim \mu_2/(\nu_2V)\ll\mu_2$, where $\nu_2$ is the density of states in the second subband. As a consequence, the system remains in the Luttinger liquid state only at temperatures $T<J$. In the intermediate temperature range, $J<T<\mu_2$, spin excitations are incoherent, and the system can be described again as an impenetrable electron gas. This phase has also been dubbed a \lq\lq spin-incoherent Luttinger liquid\rq\rq\ in the literature.\cite{Cheianov04,Fiete04,rev-Fiete}

Thus, the Hamiltonian \eqref{Band2Ham} close to the quantum phase transition at $\mu_2 = 0$ is governed by the physics of the impenetrable electron gas almost in the whole phase diagram except in the spin-coherent regime, $T \lesssim \mu_2/(\nu_2V)$, where the large spin entropy is released.

\begin{figure}
\centering
\includegraphics[width=0.15\textwidth]{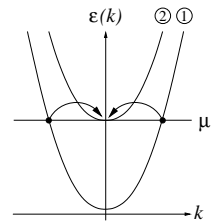}
\caption{Pair-tunneling process between the two subbands.}
\label{fig:PairTunneling}
\end{figure}

\subsubsection{Interaction between subbands}

Interactions between the subbands can be distinguished into three types. There is a repulsive density-density interaction of electrons between the subbands with amplitude $u_c > 0$, and a spin density-density interaction that is generically ferromagnetic, $u_s >0$. In addition, the amplitude $u_t$ describes pair-tunneling processes between the two subbands, see Fig.~\ref{fig:PairTunneling}. The corresponding Hamiltonian ${\cal H}_{12}$ reads
\begin{eqnarray} \label{IntHam1}
\mathcal{H}_{12} &=& \int dx\left[u_c \rho_{2}\left(\rho_{1R} + \rho_{1L} \right) - u_s \vec{S}_2 \cdot\left(\vec{S}_{1R}+\vec{S}_{1L}\right)\right]\nn\\
&& + u_t \int dx \sum_{\sigma} \Big(c^\dagger_{2 \sigma} c^\dagger_{2 \bar{\sigma}} L_{1\bar{\sigma}}R_{1\sigma} + {\rm h.c.}\Big)
\end{eqnarray}
with $\bar{\sigma} = -\sigma$. Here $\rho_{2} = \sum_{\sigma}c^{\dag}_{2\sigma}c_{2\sigma}$ and $\vec{S}_{2} = \frac{1}{2}\sum_{\sigma,\sigma'}c^{\dag}_{2\sigma}\vec{\sigma}_{\sigma,\sigma'}c_{2\sigma'}$, analogous to Eq.~\eqref{eq-rhoS}. The values for the couplings in terms of the interaction function $U$, Eq.~\eqref{Interaction2}, are given 
by 
\begin{subequations}
\begin{align}
\!\!u_c &= U_{1221}(0) \!-\! \frac{1}{2} U_{1212}(k_{F1}),
\quad
u_s = 2 U_{1212}(k_{F1}),
\\
\!\!u_t &= U_{1122}(k_{F1}),
\end{align}
\end{subequations}
and are evaluated in the next section. As in Eq.~\eqref{Band2Ham}, we neglected in \eqref{IntHam1} interaction processes involving additional spatial gradient terms.
In particular, we disregarded tunneling of electron pairs with the same spin polarization.

\begin{figure}
\includegraphics[width=0.2\textwidth]{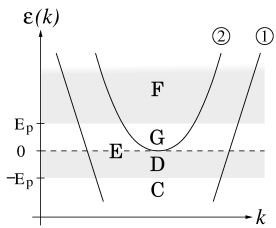}
\caption{
Energy spectrum as a function of the longitudinal momentum $k$ of a two subband quantum wire with a lower subband 1 and a higher subband 2. We consider the quantum phase transition when the chemical potential reaches the bottom of the second subband, $\mu_2 = 0$. Depending on the position of chemical potential $\mu_2$, we apply different approaches in sections
\ref{subsec:PertLargeNegMu}, \ref{subsec:ImpenetrPol}, \ref{subsec:BalentsFisher}, \ref{subsec:DenseLimit}, and \ref{subsec:G} as indicated in the figure. The energy 
scale $E_p$ is defined in Eq.~\eqref{PertEnergyScale}.
}
\label{fig:bands}
\end{figure}

The analysis of the full Hamiltonian \eqref{HamPert2} is complicated by the fact that the quantum critical point at $\mu_2 = 0$ for the interacting system has multiple dynamical scales.\cite{Zacharias09,Garst10} Whereas the Luttinger liquid has a dynamical exponent $z=1$, the spectrum of the second subband is characterized by $z=2$. The multiple scales lead to the appearance of two different types of infra-red divergences in perturbation theory. The linear spectrum of the first subband, $z=1$, yields logarithmic divergences while the quadratic spectrum of the second subband, $z=2$, yields square-root singularities. In order to understand the interplay between these two types of divergences, we start in Secs.~\ref{subsec:Pert}--\ref{subsec:ImpenetrPol} by considering the case $\mu_2<0$, where the two-particle Green function of the decoupled second (empty) subband is known exactly, see Eq.~\eqref{G2}. Then, in Sec.~\ref{subsec:BalentsFisher}, we discuss the behavior of the system at the transition, $\mu_2\approx 0$, and finally address the regime $\mu_2>0$ in Sec.~\ref{subsec:DenseLimit} and Sec.~\ref{subsec:G}. The various regimes are indicated in Fig.~\ref{fig:bands}.
 
\subsubsection{Magnitude of the coupling constants}
\label{subsub:CouplingConst}

Before turning to the analysis of the full effective low-energy Hamiltonian, let us determine the magnitude of the various coupling constants close to the phase transition. The Hamiltonian \eqref{HamPert2} contains six coupling constants. For $n a_B > 1$, the transition happens at $n x_0 \sim 1$, i.e., $k_{F1}^{-1}\sim x_0$. Both these length scales are smaller than the distance to the gate, $k_{F1}^{-1}\sim x_0 \ll d$. Due to the long-range nature of interactions, the interaction constants corresponding to processes with zero momentum transfer are enhanced by a large logarithm. Furthermore, the dependence of $U_{n_1n_2n_3n_4}$ on the subband indices becomes negligible as $d\gg x_0$. Thus,
\begin{subequations}
\label{CouplingsHighDensity}
\begin{align}
g_{1s}, u_s, u_t  \sim&\, \frac{1}{\nu_1 n a_B}
\\
g_{1c}, V, u_c \sim&\, \frac{1}{\nu_1 n a_B} \ln \frac{d}{x_0}.
\end{align}
\end{subequations}
Note that, when the Wigner crystal regime is approached, $n a_B \to 1$, the dimensionless couplings $\nu_1g$ become of order one (apart from the logarithmic enhancement\cite{Fogler05}), as expected.

\subsection{Perturbative analysis in the dilute limit $\mu_2 < 0$}
\label{subsec:Pert}

If the chemical potential of the second subband is negative, $\mu_2 < 0$, the ground state in the absence of inter-subband interactions consists of a Luttinger liquid in the first subband, Eq.~\eqref{Band1Ham1}, and an empty second subband. In the following, we consider the perturbative renormalization of the intra- and inter-subband interactions. The perturbation theory in the inter-subband interactions is particularly simple because particle-hole polarizations of the second subband vanish exactly at $T=0$. This allows to study the approach to the quantum phase transition for $\mu_2 \to 0^-$.

Perturbation theory encounters various singular corrections that diverge in the low-energy limit for $\mu_2 \to 0^-$. Below, we list these singular correction to the various couplings given at some finite energy scale $\omega$. As mentioned above, due to the multiple dynamics of the critical point, different types of divergences, logarithmic and square-root singularities, can be distinguished. The type of divergence depends on the intermediate state. A square-root divergence requires that in the intermediate state both particles are in the second subband. If at least one particle in the intermediate state is in the first subband, a logarithmic divergence is obtained. Here one can further distinguish between logarithmic divergences that are cut off by the chemical potential $\mu_2$, if one of the particles in the intermediate state is in the second subband, and those that are not, when both particles are in the first subband. 

\begin{figure}
(a)\includegraphics[width=0.14\textwidth]{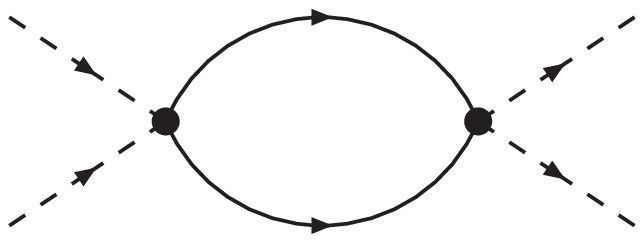}
\,\,
(b)\includegraphics[width=0.11\textwidth]{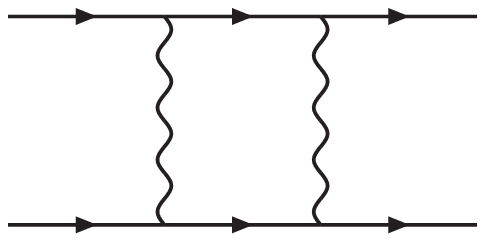}
\,\,
(c)\includegraphics[width=0.11\textwidth]{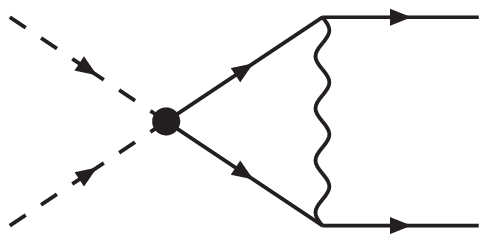}
\caption{Vertex corrections that involve the two-particle Green function of the second subband and diverge as $1/\sqrt{\varepsilon}$ with vanishing energy $\varepsilon$: correction to (a) the intra-subband interactions $g_{1x}$, $x=c,s$; (b) the intra-subband interaction $V$, and (c) the pair-tunneling $u_t$. The dashed/solid lines represents the propagator in the first/second subband, the dot is the pair-tunneling amplitude $u_t$, and the wiggly line is the intra-subband interaction $V$. 
}
\label{fig:PertVertexCorr}
\end{figure}

The perturbative corrections to the intra-subband interactions in the occupied first subband are given by
\begin{subequations} \label{SingCorrBand1}
\begin{align}
\delta g_{1c} =& \frac{1}{2}u_t^2 \mathcal{D}^{(0)}_2(0,\omega),\\
\delta g_{1s} =& -\frac{g_{1s}^2}{2\pi v_{F1}}\ln \frac{E_0}{|\omega|} + 2u_t^2\mathcal{D}^{(0)}_2(0,\omega), \label{SingCorrBand1-s}
\end{align}
\end{subequations}
where $E_0$ is a UV cutoff on the order of the Fermi energy of the occupied first subband, $E_0 \sim E_{F1}$.
Apart from the standard logarithmic corrections,\cite{GiamarchiBook} there are additional singular corrections due to the pair-tunneling $u_t$ between the subbands. The corresponding diagram is shown in Fig.~\ref{fig:PertVertexCorr}(a). These singular corrections yield square-root divergences as they involve the bare two-particle Green function $\mathcal{D}^{(0)}_2(0,\omega)$, see Eq.~\eqref{G2Bare1}, that behaves as $1/\sqrt{\omega + 2\mu_2}$.

The interactions of particles within the empty subband obtains the singular correction
\begin{align}\label{SingCorrBand2}
\delta V = V^2 \mathcal{D}^{(0)}_2(0,\omega) - \frac{u_t^2}{\pi v_{F1}} \ln \frac{E_0}{|\omega|}. 
\end{align}
The first term on the right hand side corresponds to the first term in the expansion of the two-particle T-matrix \eqref{Tmatrix} in powers of $V$, see Fig.~\ref{fig:PertVertexCorr}(b). The second term due to pair-tunneling to the first subband lowers the effective interaction. It is represented by a diagram similar to Fig.~\ref{fig:PertVertexCorr}(a), but with the dashed and solid lines interchanged.

Finally, the corrections to the inter-subband interactions read
\begin{subequations} \label{SingCorrBand12}
\begin{align}
\delta u_c =&\, \frac{u_t^2}{2\pi v_{F1}} \ln \frac{E_0}{|\omega +\mu_2|},
\\
\delta u_s =&\, -\frac{u_s^2}{2\pi v_{F1}} \ln \frac{E_0}{|\omega + \mu_2|}, \label{SingCorrBand12-s}
\\ \label{SingCorrBand12-ut}
\delta u_t =&\, V u_t \mathcal{D}^{(0)}_2(0,\omega) 
- \frac{(g_{1c} + \frac{3}{4}g_{1s})u_t}{2\pi v_{F1}} \ln \frac{E_0}{|\omega|}  
\nn\\&
+ \frac{2 u_t u_c}{\pi v_{F1}} 
\ln \frac{E_0}{|\omega + \mu_2|}.
\end{align}
\end{subequations}
The two-particle Green function of the second subband enters also here, namely in the correction to the pair-tunneling vertex, see Fig.~\ref{fig:PertVertexCorr}(c).

The singular vertex corrections listed above are either regularized by a finite energy $\omega$ or a finite chemical potential $\mu_2$.  
We will first focus on the singular corrections that survive in the limit of large negative $\mu_2$.

\subsection{Perturbative regime at large negative $\mu_2$: dilute weakly interacting Fermi gas in the second subband 
}
\label{subsec:PertLargeNegMu}

For large negative chemical potential of the second subband, the vertex corrections involving the two-particle Green function of Fig.~\ref{fig:PertVertexCorr} can be treated perturbatively. The relevant energy scale is found by estimating the value of the square-root corrections in the limit $\omega\to0$ and requiring $\delta x/x<1$, where $x=g_{1c},g_{1s},V,u_t$. Using ${\cal D}_2^{(0)}\propto\sqrt{m/\mu_2}$, one obtains $|\mu_2| 
\gg E_p\equiv{\rm max}\{mu_t^4/g_{1c}^2, mu_t^4/g_{1s}^2, mV^2\}$. Thus, for our model parameters, Eqs.~\eqref{CouplingsHighDensity}, the energy scale $E_p$ is determined by the intra-subband scattering $V$,
\begin{align} \label{PertEnergyScale}
E_p =
m V^2. 
\end{align}
Similarly, for $|\mu_2| \gg E_p$, the logarithmic corrections involving the chemical potential $\mu_2$ in their argument will reduce to small perturbative corrections only. The remaining logarithmic corrections to $g_{1s},V,u_t$ that are singular in the limit $\omega \to 0$ derive from exciting the Luttinger liquid and can be summed up with the help of the conventional RG approach. After integrating out modes within an energy shell $(E_0/b, E_0)$ with the scaling parameter $b > 1$, we obtain the RG equations
\begin{subequations}
\label{eq-mu2large}
\begin{align}
\frac{\partial g_{1s}}{\partial \ln b} =& -\frac{g_{1s}^2}{2\pi v_{F1}},\\
\frac{\partial V}{\partial \ln b} =&- \frac{u_t^2}{\pi v_{F1}},\\
\frac{\partial u_t}{\partial \ln b} =&- \frac{(g_{1c}+3g_{1s}/4) u_t}{2\pi v_{F1}}.
\end{align}
\end{subequations}
The solution of these equations is straightforward. The first equation is the standard RG equation for the spin mode of a Luttinger liquid.\cite{GiamarchiBook} The coupling
$g_{1s}$ is marginally irrelevant so that it vanishes logarithmically, 
$g_{1s}(b)/(\pi v_{F1}) \propto 1/\ln b$, for vanishing energies, $b \to \infty$. 
The last two equations describe the renormalizations of the interaction $V$ in the second subband and the pair-tunneling $u_t$ due to the interaction with the Luttinger liquid. 

Although the latter two interactions, $V$ and $u_t$, involve high-energy excitations to the second subband that is far away in energy ($|\mu_2| \gg E_p$), it is nevertheless important to consider them in order to check for consistency. In particular, if $|V|$ flows to large values, the condition $|\mu_2| \gg E_p = m V^2$ may be violated during the RG flow. The pair-tunneling $u_t$ leads to a reduction of the interaction $V$ and could potentially drive it to large negative values. A change of sign of $V$ from repulsive to attractive is accompanied by the appearance of a two-particle bound state at energies below the bottom of the second subband. In that case, the transition at $\mu_2=0$ could be preempted by the condensation of bound pairs. 

However, it turns out that the RG flow of the interaction $V$ is rather short. The pair-tunneling $u_t$ which drives the flow of $V$ suffers an orthogonality catastrophe, i.e., it is suppressed by the interactions in the Luttinger liquid. Thus, $u_t$ is irrelevant and flows to zero with a scaling dimension that approaches $g_{1c}/(2\pi v_{F1})$ in the low-energy limit, $b \to \infty$. It is straightforward to solve Eqs.~\eqref{eq-mu2large} for the effective coupling $V_{\rm eff}$ at the lowest energies. Namely,
\begin{align} \label{EffScatteringV}
V_{\rm eff} = V - \frac{u_t^2}{g_{1c}} \phi\big(\frac{g_{1s}}{g_{1c}}\big),
\end{align}
where the function $\phi$ has the limits $\phi(0) = 1$ and $\phi(x) \approx 4/x$ for $x\to \infty$. Using the initial values of the interaction constants, Eq.~\eqref{CouplingsHighDensity}, one obtains $\delta V/V\sim1/(\ln d/x_0)^2$, i.e., 
for $d \gg x_0$ the correction is logarithmically small. Thus, the effective interaction $V_{\rm eff}$ remains repulsive and in the perturbative regime, and, consequently, the condition $|\mu_2| \gg m V^2_{\rm eff}$ still holds.
 
To summarize, in the limit of large negative chemical potential, $|\mu_2| \gg E_p$, the ground state of the system is not affected by the inter-subband interactions which remain perturbative. The first subband is a Luttinger liquid while the second subband corresponds to a weakly-interacting dilute Fermi gas whose population is exponentially small in ${\mu_2/T}$ for $|\mu_2| \gg T$.

\subsection{Quantum phase transition at $\mu_2 = 0$: Lifshitz transition of impenetrable polarons}
\label{subsec:ImpenetrPol}

If the negative chemical potential $\mu_2<0$ increases and passes the threshold $|\mu_2| \sim E_p$, see Eq.~\eqref{PertEnergyScale}, the square-root singularities of the two-particle Green function $\mathcal{D}_2^{(0)}$ develop in the perturbative corrections to the couplings $g_{1c},g_{1s},V$, and $u_t$, see Eqs.~(\ref{SingCorrBand1},\ref{SingCorrBand2},\ref{SingCorrBand12-ut}) and Fig.~\ref{fig:PertVertexCorr}. In the following, we consider the limit of small negative chemical potential, $|\mu_2| \ll E_p$, and, in particular, the approach to the quantum phase transition $\mu_2 \to 0^-$.

\subsubsection{Formation of an impenetrable electron gas in the second subband}
\label{subsec:ImpenetrPol1}

In the limit $|\mu_2| \ll E_p$, the inverse square-root singularities that appear in the perturbation theory become of order one for energies $\omega \sim E_p$, while the logarithmic corrections are still small. So it is permissible to consider first the stronger inverse square-root singularities deriving from the two-particle Green function of the second subband. In order to identify the class of most divergent diagrams, we have to consider processes that contain as many two-particle propagators $\mathcal{D}_2^{(0)}$ as possible. These processes are associated with the repeated scattering of particles in the second subband. The summation of this class of diagrams corresponds to the formation of an impenetrable electrons gas in the second subband as discussed in Sec.~\ref{sssec-H2}. As a consequence the bare two-particle propagator $\mathcal{D}_2^{(0)}$ appearing in Eqs.~(\ref{SingCorrBand1},\ref{SingCorrBand2},\ref{SingCorrBand12-ut}) has to be replaced by the full two-particle propagator $\mathcal{D}_2$.

Consider first the renormalization of the intra-subband interaction $V$ in the second subband. The summation of repeated two-particle scattering processes promotes the interaction $V$ to the T-matrix, $V \to V_{\rm eff} = V+V^2\mathcal{D}_2(k,\omega)=\mathcal{T}(k,\omega)$, see Eq.~\eqref{Tmatrix} and Fig.~\ref{fig2}. As discussed in Sec.~\ref{sssec-H2}, the effective interaction thus becomes explicitly energy dependent and, as the distance $\varepsilon = \omega - k^2/(4m) + 2 \mu_2$ to the two-particle mass shell further decreases, $|\varepsilon| < E_p$, it assumes the universal form $\mathcal{T}(k,\omega) \approx -1/\mathcal{D}^{(0)}_2(k,\omega)$, which is characteristic for an impenetrable electron gas.

\begin{figure}
(a)\parbox[b]{0.4\textwidth}{\includegraphics[width=0.4\textwidth]{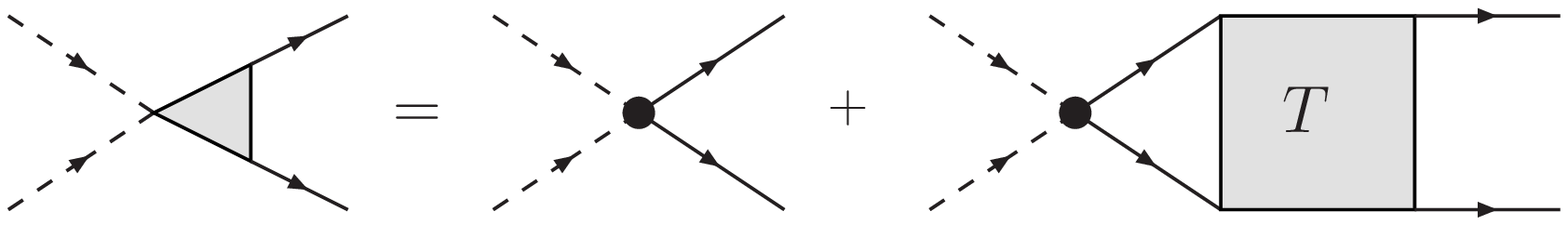}}\\[0.8em]
(b)\,\,\parbox[b]{0.4\textwidth}{\includegraphics[width=0.4\textwidth]{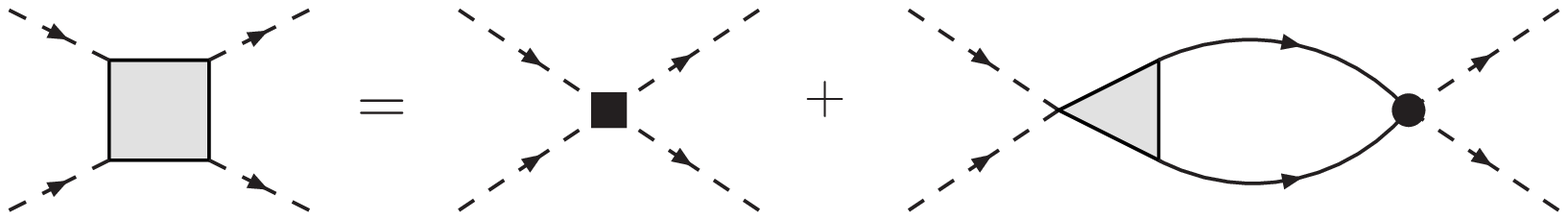}}
\caption{(a) Renormalization of the pair-tunneling vertex $u_t$. Two electrons of the first subband (dashed lines) are transfered to the second subband (solid lines). In addition to the bare pair-tunneling (black dot), the vertex receives renormalizations from the repeated scattering of electrons in the second subband that is captured by the T-matrix, see Eq.~\eqref{Tmatrix} and Fig.~\ref{fig2}. (b) Renormalization of the intra-subband interactions $g_{1x}$ with $x=c,s$ within the first subband. The black square represents the bare interaction; the renormalization is due to pair-tunneling into the second subband with the renormalized pair-tunneling vertex defined in (a). 
}
\label{fig3}
\end{figure}

The formation of an impenetrable electron gas in the second subband has the consequence that the pair-tunneling becomes inefficient because the probability to find two electrons at the same position in space is strongly suppressed. The singular vertex corrections for the pair-tunneling are shown in Fig.~\ref{fig3}(a). In the limit $\varepsilon\to0$, summation of these diagrams gives
\begin{equation} \label{EffPairTunneling}
u_t(\varepsilon) = 
u_t + V u_t \mathcal{D}_{2}(k,\omega)
\to -\frac{u_t}{V \mathcal{D}^{(0)}_2(k,\omega)}.
\end{equation}
Thus, the effective pair-tunneling vanishes as 
\begin{align} \label{RenPairTunneling}
u_t(\varepsilon) \sim u_t\sqrt{\frac\varepsilon {mV^2}}\quad {\rm for}\quad \varepsilon \to 0,
\end{align}
i.e., pair-tunneling is suppressed by a factor $\sqrt{\varepsilon/E_p}$. 

Furthermore, the renormalizations of the Luttinger liquid interactions due to pair-tunneling, Eqs.~\eqref{SingCorrBand1}, are now regularized by the strong interactions in the second subband, see Fig.~\ref{fig3}(b). Replacing $\mathcal{D}_2^{(0)}$ by $\mathcal{D}_2$ in Eqs.~\eqref{SingCorrBand1} and taking the limit $\varepsilon\to0$, one obtains
\begin{align} \label{EffInteractionBand1}
g_{1x}^{{\rm eff}} = g_{1x} - C_x\frac{u_t^2}{V},
\end{align}
where $x=c,s$ and the constants $C_c=1/2, C_s=2$.

Thus, the pair-tunneling $u_t$ leads to a negative correction to the Luttinger liquid interactions. If $u_t$ is sufficiently large, it may lead to a sign change of the interaction constants and, thus, induce attractive (anti-ferromagnetic) interactions in the Luttinger liquid. In particular, the smaller effective coupling $g_{1s}^{\rm eff}$, that controls the spin sector, turns negative if
\begin{align} \label{SpinGap}
\frac{u_t^2}{V g_{1s}} \gtrsim \mathcal{O}(1).
\end{align}
If the effective coupling were driven negative, $g_{1s}^{\rm eff}<0$, the interactions between the residual low-energy degrees of freedom would drive the interaction to strong coupling and generate a spin gap in
the first subband,\cite{GiamarchiBook} see also discussion in the next section. 
However, for the parameters of Eqs.~\eqref{CouplingsHighDensity} we have $u_t^2/(V g_{1s})\sim 1/\ln (d/x_0)$ so that the criterion \eqref{SpinGap} is never fulfilled for $d \gg x_0$. Thus, the Luttinger liquid in the first subband remains gapless.

So far we considered the effect of repeated scattering of particles in the second subband yielding the most singular corrections.
We found that, as a consequence, an impenetrable electron gas forms in the second subband at energies $\omega \sim E_p$, and the pair-tunneling between the subbands becomes ineffective due to the formation of nodes in the two-particle wavefunctions. However, the remaining interaction still leads to interesting physics that develops for even smaller energies, $\omega \ll E_p$, as we show below.

\subsubsection{Polaron formation at lowest energies
} 
\label{subsubsec:QPT}

Due to the formation of an impenetrable electron gas in the second subband, the pair-tunneling is suppressed by a factor $\sqrt{\omega/E_p}$, see Eq.~\eqref{RenPairTunneling}, and consequently becomes ineffective for energies $\omega \lesssim E_p$. At the lowest energies, $\omega \ll E_p$, the pair-tunneling may therefore be neglected. However, logarithmic divergences in Eqs.~(\ref{SingCorrBand1-s},\ref{SingCorrBand12-s}) survive and lead to a residual flow of the remaining vertices. A perturbative treatment of the vertex corrections at energies $\omega < E_p$ yields the one-loop RG equations
\begin{subequations}
\label{RGLowestEnergies}
\begin{align} 
\frac{\partial g_{1s}}{\partial \ln b} =& -\frac{g_{1s}^2}{2\pi v_{F1}},\label{eq-RG_LLs}
\\
\frac{\partial u_s}{\partial \ln b} =& -\frac{u_s^2}{2\pi v_{F1}}.\label{eq-RG_inter-s}
\end{align}
\end{subequations}
Whereas the first equation derives from processes in the Luttinger liquid of the first subband only, the flow of $u_s$ 
is caused by the virtual excitation of a particle to the empty second subband. Contrary to the pair-tunneling process, there is only a single particle involved which -- as the second subband is empty -- cannot repeatedly scatter from other particles. As a result, $u_s$ is not suppressed by the formation of an impenetrable electron gas in the second subband even for $\omega < E_p$. 

To solve the RG equations, it is sufficient to consider the logarithmic renormalizations arising at the lowest energies, i.e., the effective cut-off can be chosen as $E_0 \lesssim E_p$. Consequently, the initial values for the RG flow are given by the values of the coupling constants at the energy scale $E_p$, namely the effective coupling $g^{\rm eff}_{1s}$ of Eq.~\eqref{EffInteractionBand1} and the bare inter-subband coupling $u_s$.

Eq.~\eqref{eq-RG_LLs} is the standard RG equation for the Luttinger liquid. Depending on the sign of 
$g_{1s}^{\rm eff}$, the flow is either to weak or strong coupling. As discussed in the context of Eq.~\eqref{EffInteractionBand1}, for $d/x_0 \gg 1$ the coupling
$g_{1s}^{\rm eff}$ is positive, corresponding to ferromagnetic spin-spin interactions, and thus flows to weak coupling.

Eq.~\eqref{eq-RG_inter-s} for the inter-subband spin-spin interaction has the same form as Eq.~\eqref{eq-RG_LLs}. Thus, the (positive) coupling constant $u_s(\omega)$ also decreases logarithmically with decreasing $\omega$,
\begin{align} \label{FlowIntrabanSpin}
u_s(\omega) = \frac{u_s}{1+\frac{u_s}{2\pi v_{F1}} \ln \frac{E_p}{{\rm max}\{\omega,|\mu_2|\}}},
\end{align}
i.e., the ferromagnetic inter-subband spin-spin interaction is marginally irrelevant. Contrary to the intra-subband spin-spin interaction, here the flow is stopped for a finite (negative) chemical potential $|\mu_2| \ll E_p$. 

\begin{figure}
\includegraphics[width=0.2\textwidth]{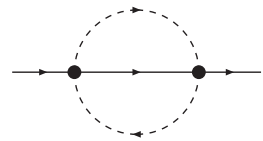} 
\caption{Self-energy for electrons in the second subband. The dot represents either the charge or spin inter-subband interaction vertex, $u_c$ or $u_s$, respectively. 
}
\label{fig:Sigma}
\end{figure}

Even though the pair-tunneling is suppressed by the strong interactions in the second subband, the low-energy limit of the system does not correspond to two decoupled subbands, namely a Luttinger liquid and an impenetrable electron gas. The inter-subband charge and spin density interactions, $u_c$ and $u_s$, respectively, lead to a polaron effect.\cite{Balents00} The particles of the second subband occupy states close to the band bottom, see Fig.~\ref{fig:bands}, and as a consequence, they propagate very slowly: their velocity vanishes close to the band bottom. By comparison, the density fluctuations in the filled first subband are rather fast, and the density can adiabatically adjust to screen the particles in the second subband. This adjustment is reflected in logarithmically singular corrections to the residue $Z$ of the fermionic single-particle Green function in the second subband, ${\cal G}_2^{-1}(\omega,k) = \omega - k^2/(2m) + \mu_2 - \Sigma(\omega,k)$, that only appears in two-loop order. Evaluating the self-energy diagram shown in Fig.~\ref{fig:Sigma}, one obtains 
\begin{align} \label{SelfEnergyPolaron}
\Sigma(\omega,k) = \left(\omega - \frac{k^2}{2m} + \mu_2\right) 
\frac{2u_c^2+\frac38u_s^2}{(2\pi v_{F1})^2}
\\\nn \times \ln \frac{E_0}{\omega - \frac{k^2}{2m} + \mu_2}.
\end{align}
Note that Fig.~\ref{fig:Sigma} represents the only important two-loop self-energy diagram for $\mu_2<0$ and $T=0$. In particular, contributions arising from intra-subband scattering $V$ are exponentially suppressed in $\mu_2/T$, see Ref.~\onlinecite{Goehmann98-2}. The logarithmically singular self-energy $\Sigma$ results in an RG flow for the residue $Z$ giving rise to an anomalous dimension for the electrons in the empty second subband. The flow can be integrated after also taking into account the two-loop vertex corrections.\cite{Balents00} Note that not only charge but also spin excitations (absent in the spin polarized case considered in Ref.~\onlinecite{Balents00}) contribute to screening and consequently to the logarithmic singularity in Eq.~\eqref{SelfEnergyPolaron}. 

\subsubsection{Universality class of the quantum phase transition}

After all, we can identify the quantum phase transition occurring at $\mu_2=0$ as a Lifshitz transition corresponding to the filling of an empty subband as a function of the chemical potential $\mu_2$. The electrons that fill this empty subband are, however, strongly interacting, characterized by unitary scattering.  Furthermore, each of these electrons is screened by the density and spin excitations of the Luttinger liquid in the filled first subband. As a result, the quantum phase transition is a Lifshitz transition of impenetrable polarons.

In the following, we discuss a regularization scheme for the square-root singularities that allows for an unbiased renormalization group analysis and confirms the physical picture for the quantum phase transition developed above.

\subsection{Regularization of the $z=2$ singularities
at the quantum critical point $\mu_2 = 0$}
\label{subsec:BalentsFisher}

The two subband Hamiltonian, Eq.~\eqref{HamPert2}, has previously been considered by Balents and Fisher in Ref.~\onlinecite{Balents96} in the context of the two-chain Hubbard model. In order to deal with the logarithmic and square-root singularities encountered in perturbation theory, see section \ref{subsec:Pert}, they considered a generalized dispersion for the electrons in the second subband, $\varepsilon(k) = |k|^{1+\epsilon} v^{1-\epsilon} /(2 m)^\epsilon$, where $v$ is an artificial parameter with the dimension of velocity. The physical quadratic dispersion of the Hamiltonian \eqref{Band2Ham} is recovered for $\epsilon = 1$ whereas $\epsilon=0$ corresponds to a linear spectrum. Treating $\epsilon$ as a small parameter the square-root singularities are regularized and RG equations for the Hamiltonian close to criticality, $\mu_2 \approx 0$, can be derived.  In lowest order in $\epsilon$, one obtains\cite{Balents96} 
\begin{subequations}
\label{BF-RG}
\begin{align}
\label{BF-Flow-g1c}
\frac{\partial g_{1c}}{\partial \ln b} &= - \frac{u_t^2}{4\pi v},
\\\label{BF-Flow-g1s}
\frac{\partial g_{1s}}{\partial \ln b} &= - \frac{g_{1s}^2}{2\pi v_{F1}} - \frac{u_t^2}{\pi v},
\\\label{BF-Flow-V}
\frac{\partial V}{\partial \ln b} &= \epsilon V - \frac{V^2}{2\pi v} - \frac{u_t^2}{\pi v_{F1}},
\\
\frac{\partial u_c}{\partial \ln b} &= \frac{u_t^2}{2\pi (v_{F1}+v)},
\\
\frac{\partial u_s}{\partial \ln b} &= - \frac{u_s^2}{2\pi (v_{F1}+v)},
\\\label{BF-Flow-ut}
\frac{\partial u_t}{\partial \ln b} &= \left(\frac{\epsilon}{2}- \frac{V}{2\pi v}  + 
\frac{2 u_c}{\pi (v_{F1}+v)}  - \frac{g_{1c} + \frac34g_{1s}}{2\pi v_{F1}}\right)u_t.
\end{align}
\end{subequations}
Here the RG equations are derived under the assumption that all dimensionless coupling constants are much smaller than $\epsilon$.

The parameter $\epsilon$ appears in the equations for the intra-subband interaction in the second subband, $V$, and the pair tunneling, $u_t$, i.e., the coupling constants that acquire corrections due to the interaction $V$. Both $V$ and $u_t$ initially have a positive scaling dimension, $\epsilon$ and $\epsilon/2$, respectively, i.e., they are both relevant. However, the scaling dimension of $V$ is larger and, thus, the intra-subband scattering $V$ is the most relevant perturbation. Initially, $V$ therefore grows exponentially as a function of $\ln b$. As a consequence, the intra-subband scattering quickly reaches an attractive fixed point $V/(2\pi v) \to \epsilon$. This fixed point can be identified with the impenetrable electron gas within the second subband.  

The pair-tunneling initially increases as well but with the smaller scaling dimension $\epsilon/2$. As $V$ approaches the strong-coupling fixed-point, the scaling dimension of the pair tunneling changes from relevant to irrelevant as $\epsilon/2 - V/(2\pi v) \to -\epsilon/2$, corresponding to a suppression of pair-tunneling due to the formation of nodes in the two-particle wave function, see also Eq.~\eqref{RenPairTunneling}. As a consequence, the pair-tunneling quickly flows to zero and the remaining RG flow reproduces the one given in Eqs.~\eqref{RGLowestEnergies}. 

The approximations of section~\ref{subsec:ImpenetrPol1} with the expressions \eqref{EffInteractionBand1} for the effective couplings are exactly recovered if the initial exponential flow of $V$ and $u_t$ is reduced only to the first two terms on the right hand side of Eqs.~\eqref{BF-Flow-V} and \eqref{BF-Flow-ut}. Plugging the obtained result for $u_t(b)$ into Eqs.~\eqref{BF-Flow-g1c} and \eqref{BF-Flow-g1s}, one recovers Eq.~\eqref{EffInteractionBand1} with the same coefficients $C_x$.

It is interesting that the reduction of the effective interactions \eqref{EffInteractionBand1} quantified by $C_x$ can be mainly attributed to an approximate unstable fixed-point of the RG flow \eqref{BF-RG}. The interactions $g_{1c}$ and $g_{1s}$ are substantially renormalized 
when the pair-tunneling reaches its maximal value. This occurs at scales $b \approx b_m$ when the flow of $u_t$ is approximately stationary, i.e., when the intra-subband scattering reaches half of the fixed point value, $V(b_m)/(2\pi v) = \epsilon/2$. Close to this stationary point, the flow of $u_t$ can be approximated by 
\begin{align}
u_t(b) \approx u_{t {\rm max}} e^{- \frac{\epsilon^2 \ln^2 (b/b_m)}{8}}.\label{eq-umax-flow}
\end{align}
The maximal value of the pair-tunneling, $u_{t{\rm max}}$, is obtained from the approximate RG invariant of the initial flow. Considering the coupled equations \eqref{BF-Flow-V} and \eqref{BF-Flow-ut}, and keeping only the dominant first two terms in Eq.~\eqref{BF-Flow-ut}, one finds the approximate RG invariant $\mathcal{I} = (\epsilon V - V^2/(2\pi v))\pi v_{F1}/u_t^2 + 2\ln u_t$. With $V(b_m)/(2\pi v) = \epsilon/2$, the maximal pair tunneling then obtains
$u_{t{\rm max}} \approx u_t \sqrt{\pi v \epsilon/(2 V)}$, i.e., the pair-tunneling close to $b_m$ is enhanced by the large factor $\sqrt{\pi v \epsilon/(2 V)}$.

Consequently, the pair-tunneling correction will dominate the renormalization of $g_{1s}$ close to $b_m$ so that we can approximate its flow by $\partial g_{1s}/\partial \ln b \approx - u_t^2/(\pi v)$ with $u_t(b)$ given by Eq.~\eqref{eq-umax-flow}. Integrating this equation one obtains an effective intra-subband interaction 
\begin{align}
g_{1s}^{\rm eff} = g_{1s} - \sqrt{\pi}\frac{u_t^2}{V}.
\label{eq-BF_g1eff}
\end{align}
The stationary point of the RG flow thus accounts already for a reduction with a prefactor $\sqrt{\pi}$ as compared to $C_s = 2$ obtained in section~\ref{subsec:ImpenetrPol1}. Needless to say that we arrive at the same conclusion concerning the stability of the Luttinger liquid as in the previous section, see the criterion of Eq.~\eqref{SpinGap}.

This is to be contrasted with the results of Ref.~\onlinecite{Balents96}, where a Hubbard ladder was considered. In that case, all interactions are of the same order, namely on the order of the Hubbard interaction $U$. Thus, for Hubbard initial conditions the criterion \eqref{SpinGap} is fulfilled, and a spin gap can be generically expected in agreement with Ref.~\onlinecite{Balents96}. For the quantum wire with screened Coulomb interaction, however, the system remains gapless.

\subsection{Dense limit of large positive chemical potential, $\mu_2 > E_p$}
\label{subsec:DenseLimit}

The square-root singularities of the perturbation theory at $\mu_2=0$ can also be regularized by considering a finite positive chemical potential larger than the strong-coupling scale, $\mu_2 > E_p$, defined in Eq.~\eqref{PertEnergyScale}. For a positive chemical potential, the ground state contains a finite density of particles in the second subband. At lowest temperatures, $T \ll \mu_2$, we can then linearize the quadratic spectrum of the second subband and map its Hamiltonian $\mathcal{H}_2$, Eq.~\eqref{Band2Ham}, to a Tomonaga-Luttinger model. 

Analogous to the treatment of the first subband in Sec.~\ref{sssec-H2}, we introduce right- and left-movers, cf.~Eq.~\eqref{eq-RL}, to find the standard Luttinger Hamiltonian $\mathcal{H}_2^{\rm lin}$, cf.~Eqs.~(\ref{Band1Ham1},\ref{eq-rhoS}), with interaction constants $g_{2c},g_{2s}$. In the dense limit, $E_{F1}\gg\mu_2 \gg E_p$, the values of the coupling constants are
\begin{equation}
g_{2c} = \frac{V}{2}, \qquad
g_{2s} = 2V.
\end{equation}

In terms of the new fields, the inter-subband interactions \eqref{IntHam1} can be written as $\mathcal{H}^{\rm lin}_{12} = \mathcal{H}^{\rm lin}_{12 cs} +\mathcal{H}^{\rm lin}_{12 t}$ with
\begin{eqnarray}
\mathcal{H}^{\rm lin}_{12 cs}  & = & u_{c} \int dx\, \left(\rho_{1R}\rho_{2L} + \rho_{2R}\rho_{1L}\right) \\
\nn&& - u_{s}\int dx\,\left(\vec{S}_{1R}\cdot\vec{S}_{2L} + \vec{S}_{2R}\cdot\vec{S}_{1L}\right)
\end{eqnarray}
and
\begin{align}
\mathcal{H}^{\rm lin}_{12 t} & = \int dx\left[u_{tc}\left(T_{R}T_{L} + {\rm h.c.}
\right) - u_{ts}\left(\vec{T}_R\cdot\vec{T}_{L} + {\rm h.c.}
\right)\right],
\end{align}
where $T_{r} = \sum_{\sigma}r^{\dagger}_{1\sigma}r_{2\sigma}$ and $\vec{T}_{r} = \sum_{\sigma}r^{\dagger}_{1\sigma}\vec{\sigma}_{\sigma,\sigma'}r_{2\sigma'}/2$. 
For $\mu_2 \ll E_{F1}$, the charge and spin pair-tunneling interactions are related to the couplings defined in Eq.~(\ref{IntHam1}) by $u_{tc} = u_{t}/2$ and $u_{ts} = 2u_{t}$. 

The resulting Hamiltonian $\mathcal{H}_1 + \mathcal{H}^{\rm lin}_2 + \mathcal{H}^{\rm lin}_{12}$, with $\mathcal{H}_1$ of Eq.~\eqref{Band1Ham1}, was previously analyzed by Varma and Zawadowski in Ref.~\onlinecite{Varma85} and subsequently by other authors.~\cite{Penc90,Balents96}
Integrating out modes in an energy shell $(E_0/b,E_0)$ with $b>1$, where the cut-off here is on the order of the 
small chemical potential, $E_0 \sim \mu_2 \ll E_{F1}$, one obtains the following RG flow for the coupling constants,~\cite{Varma85}
\begin{subequations}
\label{RGFiniteDensity}
\begin{align}
\frac{\partial g_{1c}}{\partial \ln b} =& -\frac{1}{2\pi v_{F2}}\left(u^2_{tc} + \frac{3}{16}u^2_{ts}\right),\\
\frac{\partial g_{1s}}{\partial \ln b} =& -\frac{1}{\pi v_{F2}}\left(u_{tc}+\frac{1}{4}u_{ts}\right)u_{ts} -\frac{1}{2\pi v_{F1}}g^2_{1s},\\
\frac{\partial g_{2c}}{\partial \ln b} =& -\frac{1}{2\pi v_{F1}}\left(u^2_{tc} + \frac{3}{16}u^2_{ts}\right),\\
\frac{\partial g_{2s}}{\partial \ln b} =& -\frac{1}{2\pi v_{F2}}g^2_{2s} - \frac{1}{\pi v_{F1}}\left(u_{tc}+\frac{1}{4}u_{ts}\right)u_{ts},\label{RGFiniteDensity-g2s}\\
\frac{\partial u_{c}}{\partial \ln b} =& \frac{1}{\pi v_{F1}}\left(u^2_{tc} + \frac{3}{16}u^2_{ts}\right),\\
\frac{\partial u_{s}}{\partial \ln b} =& -\frac{1}{\pi v_{F1}}\left(u^2_{s} - 2\left(u_{tc} - \frac{1}{4} u_{ts}\right)u_{ts}\right),\\
\frac{\partial u_{tc}}{\partial \ln b} =& -\frac{1}{2\pi v_{F2}}\left(g_{2c}u_{tc} + \frac{3}{16}g_{2s}u_{ts}\right) \\
& - \frac{1}{2\pi v_{F1}}\left(\!\left(g_{1c}\!-\!4u_{c}\right)u_{tc}+\frac3{16}\left(g_{1s}\!-\!4u_{s}\right)u_{ts}\!\right)\!,\nonumber\\
\frac{\partial u_{ts}}{\partial \ln b} =& -\frac{1}{2\pi v_{F2}}\left(g_{2s}u_{tc} + \left(g_{2c} + \frac{1}{2}g_{2s}\right)u_{ts}\right)\nonumber\\
& - \frac{1}{2\pi v_{F1}}\left(g_{1s}-4u_{s}\right)u_{tc}\\ 
& - \frac{1}{2\pi v_{F1}}\left(g_{1c}+\frac12g_{1s}-4u_{c}+2u_{s}\right)u_{ts}. \nonumber
\end{align}
\end{subequations}
Note that this perturbative approach necessarily breaks down as the phase transition $\mu_2 \to 0$ is approached because the vanishing Fermi velocity $v_{F2} = \sqrt{2 m \mu_2}$ causes terms on the right hand side of the RG equations to diverge. The perturbative treatment is controlled as long as all the terms on the right hand side of Eqs.~\eqref{RGFiniteDensity} are smaller than the coupling constants they renormalize. For the parameters of our model, Eqs.~\eqref{CouplingsHighDensity}, the largest term is $\sim g_{2s}^2/v_{F2}$ in Eq.~\eqref{RGFiniteDensity-g2s}. Thus, we obtain the condition $g_{2s}/v_{F2}\ll1$ or, equivalently, $\mu_2\gg E_p$ with the strong-coupling energy scale $E_p = m V^2$ of Eq.~\eqref{PertEnergyScale}.

Close to the transition $\mu_2 \gtrsim E_p$, the initial RG flow is determined by the terms that are enhanced by factors of the inverse Fermi velocity, $v_{F2}^{-1}$.~\cite{Varma85} Keeping only such terms, the flow can be approximated as
\begin{subequations}
\label{InitialRGFiniteMu}
\begin{align}
\frac{\partial g_{1c}}{\partial \ln b} =& -\frac{1}{2\pi v_{F2}}\left(u^2_{tc} + \frac{3}{16}u^2_{ts}\right),\\
\frac{\partial g_{1s}}{\partial \ln b} =& -\frac{1}{\pi v_{F2}}\left(u_{tc}+\frac{1}{4}u_{ts}\right)u_{ts},\\
\frac{\partial g_{2s}}{\partial \ln b} =& -\frac{1}{2\pi v_{F2}}g^2_{2s},\\
\frac{\partial u_{tc}}{\partial \ln b} =& -\frac{1}{2\pi v_{F2}}\left(g_{2c}u_{tc} + \frac{3}{16}g_{2s}u_{ts}\right),\label{InitialRGFiniteMu-utc}\\
\frac{\partial u_{ts}}{\partial \ln b} =& -\frac{1}{2\pi v_{F2}}\left(g_{2s}u_{tc} + \left(g_{2c} + \frac{1}{2}g_{2s}\right)u_{ts}\right).\label{InitialRGFiniteMu-uts}
\end{align}
\end{subequations}
This initial flow shares similarities with the one obtained at large negative chemical potential $|\mu_2| \gg E_p$ in section \ref{subsec:PertLargeNegMu}, but with the role of the two subbands reversed: here only processes contribute where both particles in the intermediate state are in the second subband. The equation for the intra-subband spin interaction $g_{2s}$ decouples. As $g_{2s}>0$, the flow is to weak coupling. The pair-tunneling reduces the intra-subband interactions of the first subband, $g_{1c}$ and $g_{1s}$, 
making them less repulsive. This renormalization is, however, limited because the pair-tunneling again suffers an orthogonality catastrophe -- now due to the interactions in the second subband. Eqs.~(\ref{InitialRGFiniteMu-utc},\ref{InitialRGFiniteMu-uts}) show that the pair-tunneling is irrelevant with scaling dimension $g_{2c}/(2\pi v_{F2})$ and, thus, quickly decreases. Introducing linear combinations of coupling constants, Eqs.~(\ref{InitialRGFiniteMu-utc}) and (\ref{InitialRGFiniteMu-uts}) can be recast in the form $\partial u_{t\alpha}/\partial\ln b=-g_{2\alpha}u_{t\alpha}/(2\pi v_{F2})$, where $x_\alpha=x_c+\alpha x_s$ and $\alpha=-1/4,3/4$. Using the identifications with the coupling constants of the Hamiltonian \eqref{HamPert2}, one finds the initial conditions $u_{t(3/4)} = 2 u_t = 2 U_{1122}(k_{F1})$ and $u_{t(-1/4)} = 0$, so that the latter does not flow. Solving consecutively the RG equations for $g_{2s}$, $u_{t(3/4)}$, and then $g_{1s}$, the finite renormalizations of the intra-subband scattering $g_{1s}$ during this process can be written as
\begin{align}
g^{\rm eff}_{1s}= g_{1s} - \frac12\frac{(u_{tc} + \frac{3}{4}u_{ts})^2}{g_{2c}} \phi\big(\frac{g_{2s}}{g_{2c}}\big) =  
g_{1s} - 4 \phi(4) \frac{u_t^2}{V} 
\end{align}
with the same function $\phi$ as in Eq.~\eqref{EffScatteringV}. Here $\phi(4) \approx 0.34$.

As above, the value of $g^{\rm eff}_{1s}$
determines the fate of the subsequent flow governed by terms that are not enhanced by the inverse velocity: While for positive $g^{\rm eff}_{1s}$ 
the flow is towards weak coupling, run-away flow obtains for negative $g^{\rm eff}_{1s}$,
signaling the opening of gaps. The coupling changes sign if $u_t^2/( V g_{1s}) > \mathcal{O}(1)$
so that we reproduce again the stability criterion \eqref{SpinGap}.

In principle, the corrections to the RG flow due to terms that are not enhanced by $v_{F2}^{-1}$ could induce other instabilities. The pair-tunneling, that is irrelevant during the initial flow \eqref{InitialRGFiniteMu}, renormalizes the charge coupling $g_{2c}$ and could drive it attractive. This could, in turn, make the pair-tunneling relevant, resulting in an instability. However, in our case the criterion for this second scenario of run-away flow is even weaker than Eq.~\eqref{SpinGap}. Thus, we conclude that in the regime $\mu_2 > E_p$ the ground state of two Luttinger liquids in the two subbands is stable.

\subsection{Small positive chemical potential: $0<\mu_2<E_p$}
\label{subsec:G}

So far, we discussed all regimes in Fig.~\ref{fig:bands} close to the phase transition except the one for finite positive but small chemical potential, $0 < \mu_2 < E_p$. The physics in this regime is already strongly influenced by the strong-coupling fixed point of the impenetrable electron gas that controls the quantum phase transition. Whereas in the regime of small and negative $\mu_2$ of section \ref{subsec:ImpenetrPol} a theoretical description in terms of 
only two interacting particles
within the second subband was sufficient, now a finite density of strongly-interacting electrons populate the second subband which eludes a perturbative description. As the charge sector of the impenetrable electron gas corresponds to a spinless Fermi gas,\cite{Ogata90} we can only speculate that the physics within this strongly-interacting regime shares certain similarities with a system of spinless fermions.

In a two-subband system of {\it spinless} fermions, one generically finds an instability for small but finite $\mu_2$ because pair-tunneling becomes relevant and leads to the opening of a gap.\cite{Meyer07} The main reason is that, for the spinless case, the intra-subband interaction of the weakly populated second subband is of order $\mathcal{O}(v_{F2}^2)$ due to the Pauli principle, whereas pair-tunneling is only suppressed by a factor $v_{F2}$, namely $u_t= U_{1122}(k_{F1}-k_{F2})-U_{1122}(k_{F1}+k_{F2})$. As a result, the orthogonality catastrophe arising from adding particles to or removing particles from the second subband is substantially weakened: the RG flow does no longer contain terms of order $v_{F2}^{-1}$ as in Eqs.~\eqref{RGFiniteDensity} while the remaining terms make the pair-tunneling marginally relevant. The flow of the pair-tunneling to strong coupling signals the appearance of a gap due to a locking of the relative charge mode. Importantly, for spinless fermions there does not exist any energy scale $E_p$ that limits the range of validity of the perturbative RG in contrast to Eqs.~\eqref{RGFiniteDensity}.

It is not unlikely that a similar kind of relevant pair-tunneling mechanism might be at work in the spinfull case as well in the regime $0 < \mu_2 < E_p$  where the physics is already influenced by the impenetrable electron gas fixed-point, maybe giving rise to a gap in the charge sector.

\section{Summary \& Discussion}
\label{Summary and Discussion}

We analyzed the quantum phase transition in a quantum wire from a one-dimensional to a quasi-one-dimensional state assuming that electrons interact via a long-range Coulomb potential screened by a nearby gate. The model of the quantum wire was defined in section~\ref{Model}, and a mean-field phase diagram was presented in Fig.~\ref{fig1}. 

In the limit of strong interactions, $n a_B \ll 1$, the ground state is determined by the charge degrees of freedom. The charges minimize the Coulomb energy by forming a Wigner crystal. Exchange processes between electrons give rise to an exponentially small exchange coupling between the spins so that the spin sector is well approximated by the nearest-neighbor Heisenberg model. The phonons of the Wigner crystal and the spin excitations of the Heisenberg model thus account for a gapless charge and spin mode in the one-dimensional limit giving rise to a C1S1 phase in the notation of Ref.~\onlinecite{Balents96}. At the transition to a quasi-one-dimensional state, the 1D Wigner crystal deforms and splits into two rows, see section~\ref{Wigner Crystal quantum wire} and Fig.~\ref{fig:transition}(b). 
The transition 
is characterized by a (non-local) Ising order parameter. 
The exponentially small spin exchange between nearest neighbors depends sensitively on the distance between charges which results in a magnetoelastic coupling between the spin and phonon excitations of the Wigner crystal. 
We found that due to the Ising symmetry the spin exchange can only couple inefficiently to the square of the order parameter so that the universality class of the transition remains unaffected by the spin sector. As a consequence, the  analysis of the spin-polarized case also applies here,\cite{Sitte09} and the transition is an Ising transition with logarithmic corrections resulting from the coupling of the critical Ising modes to the longitudinal phonons. Although this latter coupling was found to be marginally irrelevant, its RG flow leads to an enhanced SU(2) symmetry accompanied by a suppression of the velocity of the excitations.\cite{Sitte09} Right after the transition to a deformed zigzag Wigner crystal, the resulting state possesses a single gapless longitudinal phonon and a gapless spin mode, corresponding to a C1S1 phase. If the electron density increases further, the fate of the spin sector depends on the strength of ring exchange processes.\cite{Klironomos}

In the limit of weak interactions, $n a_B \gg 1$, the transition can be analyzed in terms of a two-subband model of interacting electrons, see section~\ref{Multiband quantum wire}. In the one-dimensional limit, only the lowest subband is filled, forming a Luttinger liquid with a gapless charge and a gapless spin mode, i.e., it is a C1S1 phase. As the density increases, a second subband starts being populated, see Fig.~\ref{fig:transition}(a). 
This quantum phase transition is characterized by an interplay of charge and spin degrees of freedom and, more importantly, multiple dynamical scales.\cite{Sitte09,Zacharias09,Garst10}
The Luttinger liquid within the first filled subband is Lorentz-invariant and thus has a dynamical exponent $z=1$. The electrons in the empty second subband, on the other hand, are characterized by Galilean invariance and the dynamical exponent is $z=2$. As explicitly shown in section \ref{subsec:Pert}, the presence of two dynamical exponents is reflected in divergences in perturbation theory of different strengths. Whereas perturbative corrections involving excitations of the Luttinger liquid are accompanied by ubiquitous logarithms, the corrections involving electron-electron polarizations in the second subband are more singular and diverge as the inverse square-root of energy. Indeed, the electrons at the bottom of the second subband are strongly interacting. This is best understood by considering the dimensionless coupling $\nu_2V$, which diverges at the band bottom due to the diverging one-dimensional density of states $\nu_2$. As a result, the scattering between two electrons of different spin polarizations is unitary, and the electrons become impenetrable. We found that this emergent impenetrability in fact stabilizes the Luttinger liquid in the filled subband for the following reason. The system is able to gain energy by pair-tunneling processes of electrons with different spin polarizations from the filled first into the second empty subband. This effectively generates an attractive interaction in the Luttinger liquid that increases the tendency towards the formation of a spin gap. However, due to the emergent unitary scattering the probability to find two electrons at the same position in space decreases dramatically and pair-tunneling between the subbands becomes inefficient. We showed that for electrons interacting via a screened Coulomb potential the gapless spin excitations within the first subband survive at the transition, because the bare pair-tunneling between the subbands is weaker than the intra-subband interaction, see the criterion in Eq.~\eqref{SpinGap}. This is in contrast to the two-chain Hubbard model where both couplings are of the same order and pair-tunneling promotes the formation of bound electron pairs.\cite{Balents96}  
We showed in section~\ref{subsec:BalentsFisher} that the stability criterion of Eq.~\eqref{SpinGap}, that embodies the competition between intra-subband interaction and pair-tunneling, can be attributed to a stationary, i.e., a repulsive approximate fixed-point of the renormalization group flow.  
In addition, we found similarly as in the spinless case that the impenetrable electrons are screened by plasmon and spinon excitations in the first subband. The quantum phase transition at weak interactions is, thus, identified as a Lifshitz transition in terms of impenetrable polarons. 

We were not able to identify the phase and, in particular, the number of gapless modes for a small but finite filling of the second subband, because close to the Lifshitz transition the physics is still strongly influenced by the strong coupling fixed-point of the impenetrable electron gas that is beyond a perturbative treatment. We conjecture that -- similarly to the spinless case \cite{Meyer07} -- pair-tunneling processes might play here an important role as well, maybe reducing the number of gapless modes. We conjecture that pair-tunneling processes -- similar as in the spinless case\cite{Meyer07} -- might play here an important role as well, maybe reducing the number of gapless modes. If the density is further increased and the filling of the second subband is sufficiently large, the interactions can again be treated perturbatively and the filled two subbands form two decoupled Luttinger liquids (though not in the original basis) with two gapless charge and spin modes, i.e., a C2S2 phase. 

Another open question concerns the connection between the strong and the weak coupling limit. There should exist a multicritical point at intermediate densities that separates the Ising transition for $n a_B \ll 1$ from the Lifshitz transition encountered at weak coupling, $n a_B \gg 1$. The corresponding effective theory has not been identified so far.

\acknowledgments
Helpful discussions with K. Matveev and A. Rosch are gratefully acknowledged. T.M. and M.G. are supported by SFB 608 and the research unit 960 of the DFG.  M.D. and J.M. are supported by the U.S. Department of
Energy, Office of Science, under Contract No.~DE-FG02-07ER46424. Furthermore, M.D. thanks the IMI Program of the National Science Foundation for partial support
under Award No.~DMR-0843934.



\begin{thebibliography}{99}
\bibitem{footnote1}
The exchange interaction depends on the positions of all particles in the crystal. However, the effect of fluctuations in the positions of the surrounding particles on the exchange constant is small.~\cite{Klironomos05}
\end{thebibliography}
\end{document}